\documentclass{aa}  

\usepackage{graphicx}
\usepackage{txfonts}
\usepackage{xcolor}
\usepackage{caption}
\usepackage{subfig}
\usepackage{booktabs}
\usepackage{natbib}
\usepackage[urlcolor = black, citecolor = blue, colorlinks = true]{hyperref} 
\bibpunct{(}{)}{;}{a}{}{,}             

\newcommand{\logrhk}{$\rm log\,R^{\prime}_\mathrm{HK}$}

\newcommand{\msun}{$M_{\odot}$}
\newcommand{\rsun}{$R_{\odot}$}

\newcommand{\teff}{$T_{\rm eff}$}
\newcommand{\logg}{log\,{\it g$_\star$}}
\newcommand{\feh}{[Fe/H]}

\newcommand{\mjup}{M$_J$ }

\begin{document}


   \title{The GAPS Programme at TNG.  \\
    LIII. New insights on the peculiar XO-2 system
    \thanks{Based on observations made with the Italian \textit{Telescopio Nazionale Galileo} (TNG), operated on the island of La Palma by the INAF - \textit{Fundaci\'on Galileo Galilei} at the \textit{Roque de Los Muchachos} Observatory of the \textit{Instituto de Astrof\'isica de Canarias} (IAC).}}

   \subtitle{}

   \author{A. Ruggieri\inst{1,2}
          \and
          S. Desidera\inst{2}
          \and
          K. Biazzo\inst{3}
          \and
          M. Pinamonti\inst{4}
          \and
          F. Marzari\inst{5}
          \and
          G. Mantovan\inst{1,2} 
          \and
          A. Sozzetti\inst{4}
          \and
          A. S. Bonomo\inst{4}
          \and
          A. F. Lanza\inst{6} 
          \and
          L. Malavolta\inst{1,2}
          \and
          R. Claudi\inst{2}
          \and
          M. Damasso\inst{4}
          \and
          R. Gratton\inst{2}
          \and
          D. Nardiello\inst{2}
          \and
          S. Benatti\inst{7}
          \and
          A. Bignamini\inst{8} 
          \and
          G. Andreuzzi\inst{3,9}
          \and
          F. Borsa\inst{10}
          \and
          L. Cabona\inst{2}
          \and
          C. Knapic\inst{8}
          \and
          E. Molinari\inst{10}
          \and
          L. Pino\inst{11}
          \and
          T. Zingales\inst{1}
          }

   \institute{Department of Physics and Astronomy, University of Padova,
              Vicolo dell'Osservatorio 2, I - 35122, Padova, Italy\\ 
              \email{alessandro.ruggieri.1@phd.unipd.it}
         \and
             INAF, Astrophysical Observatory of Padova, Vicolo dell'Osservatorio 5, I - 35122, Padova, Italy 
         \and
             INAF, Astronomical Observatory of Rome, Via Frascati 33, I - 00178, Monte Porzio Catone (RM), Italy  
        \and
            INAF, Astrophysical Observatory of Turin, Via Osservatorio 20, I - 10025, Pino Torinese (TO), Italy  
        \and
            Department of Physics and Astronomy, University of Padova, via Marzolo 8, I - 35131, Padova, Italy  
        \and   
            INAF, Astrophysical Observatory of Catania, Via S. Sofia 78, I - 95123 Catania, Italy 
        \and
            INAF, Astronomical Observatory of Palermo, Piazza del Parlamento 1, I - 90134, Palermo, Italy  
        \and
            INAF, Astronomical Observatory of Trieste, via Tiepolo 11, 
            I - 34143, Trieste, Italy  
        \and
            Fundación Galileo Galilei - INAF, Rambla José Ana Fernández Pérez 7, 38712, Breña Baja, Tenerife, Spain  
        \and
            INAF, Astronomical Observatory of Brera, Via E. Bianchi 46, I - 23807 Merate (LC), Italy  
        \and
            INAF, Astrophysical Observatory of Arcetri, Largo Enrico Fermi 5, I - 50125, Firenze, Italy  
        }

   \date{Received ; accepted }

 
  \abstract
   {Planets in binary systems are a fascinating and yet poorly understood phenomenon. Since there are only a few known large-separation systems in which both components host planets, characterizing them is a key target for planetary science.}
   {In this paper, we aim to carry out an exhaustive analysis of the interesting XO-2 system, where one component (XO-2N) appears to be a system with only one planet, while the other (XO-2S) has at least three planets.}
   {Over the last 9 years, we have collected 39 spectra of XO-2N and 106 spectra of XO-2S with the High Accuracy Radial velocity Planet Searcher for the Northern emisphere (HARPS-N) in the framework of the Global Architecture of Planetary Systems (GAPS) project, from which we derived precise radial velocity (RV) and activity indicator measurements. Additional spectroscopic data from the High Resolution Echelle Spectrometer (HIRES) and from the High Dispersion Spectrograph (HDS), and the older HARPS-N data presented in previous papers, have also been used to increase the total time span. We also used photometric data from TESS to search for potential transits that have not been detected yet. For our analysis, we mainly used PyORBIT, an advanced Python tool for the Bayesian analysis of RVs, activity indicators, and light curves.}
   {We found evidence for an additional long-period planet around XO-2S and characterized the activity cycle likely responsible for the long-term RV trend noticed for XO-2N. The new candidate is an example of a Jovian analog with $m\sin i \sim 3.7$ \mjup, $a \sim 5.5$ au, and $e = 0.09$. We also analyzed the stability and detection limits to get some hints about the possible presence of additional planets. Our results show that the planetary system of XO-2S is at least one order of magnitude more massive than that of XO-2N. 
   The implications of these findings for the interpretation of the previously known abundance difference between components are also discussed. }
   {}


   \maketitle
%

\section{Introduction}
In recent years, the number of discovered exoplanets has constantly increased thanks to the great effort of the astronomical community and technological advancements. In this context, a research field that has not greatly advanced yet is that of planetary systems around binary stars. These are typically more difficult to study but they can yield interesting information regarding the formation and evolution processes. This is important because binaries represent about one-third of all the stars in our Galaxy \citep[e.g.,][]{raghavan2010}, and therefore a complete comprehension of planetary systems requires the study of these objects. Planets orbiting around one component of a binary system are called S types, and several works have tried to determine how and when the stellar companion can affect them. As reported by \cite{eggenberger2010}, S-type planets do indeed display different properties than those around a single star. This is mainly driven by the dynamical interactions with the companion, in particular for binaries closer than 100 au. The difficulty in detecting and characterizing planets in close binaries often leads to their exclusion from planet search surveys. On the other hand, wide binaries are more favorable targets and can be taken into consideration for a proper analysis of the population of planets in binary systems.

Binary systems in which both stars host planets are of particular interest as it is possible to carry out comparative studies since the two stars likely formed from the same cloud (i.e., the same chemical composition and age) and experienced similar dynamical perturbations. Very few such cases of double-planet hosts are currently known: HD 20781 and HD 20782 (\citealt{jones2006} and \citealt{udry2019}), WASP-94 \citep{neveu2014}, Kepler-132 \citep{lissauer2014}, and XO-2, the first binary system for which both components were found to host planets, which is the target of this paper.
In this context, the XO-2 system is of great interest as it presents unique features that make it an important target. In particular, it includes two widely separated stars XO-2S and XO-2N (separation 31", projected separation $\sim$ 4600 au) that are almost identical to each other and to the Sun, except that they both have metallicities more than twice that of the Sun.
\cite{burke2007} discovered a transiting hot Jupiter around XO-2N, while
\cite{desidera2014} identified two planetary companions around XO-2S, a warm Saturn-like and a temperate Jupiter-like one, through radial velocity (RV) monitoring. Long-term RV trends were seen for both components \citep{damasso2015} and the one of XO-2N may be due to an activity cycle.
Further characterization of the transiting planet showed a well-aligned configuration through the Rossiter-McLaughlin effect (\citealt{narita2011} and \citealt{damasso2015}). The photometric variability and the stellar activity were studied by \cite{zellem2015} and \cite{damasso2015}, with tentative measurements of the rotation period of XO-2N and an indication that it is slightly more active, possibly because of interactions with its very close-in planet. Independent studies on the chemical abundances of the two components showed that XO-2N is also slightly more metal rich than XO-2S (\citealt{teske2015}, \citealt{biazzo2015}, and \citealt{ramirez2015}) with indications of a dependence of the abundance difference from the condensation temperature. This is interesting because visual binaries are expected to form from the same cloud, and they should thus have the same composition. In the first case, the "original" composition is that of the S component, and the N star has been enriched because the inward migration of planet b caused a potential companion to collide with the star, releasing rocky material in the stellar atmosphere. In the other scenario, XO-2S is instead depleted of metals with respect to the original composition because its heavy elements are locked into the planetary cores and therefore the star accreted (relatively) metal-depleted gas in the final stages of formation. 
In order to shed further light on this system of special interest, the first step is to fully characterize the planetary systems of both stars, in particular by analyzing the previously mentioned long-term RV trends and understanding their origin. This was accomplished with the continuation of the RV monitoring of both components for an additional 9 years with respect to \cite{damasso2015}. We also took advantage of the available TESS data to look for transits of XO-2S b and c, and for stellar characterization.
Sect. \ref{sec:obs} describes the new observations, the archive data, and the adopted data analysis procedures. In Sect. \ref{sec:starparam}, we briefly describe the stellar parameters adopted in our analysis. In Sect. \ref{sec:xo2s} and \ref{sec:xo2n} we present the results for XO-2S and XO-2N, respectively. In Sect. \ref{sec:discussion}, the implications of our findings are discussed.
In Sect. \ref{sec:conclusion} we summarize our conclusions.

\section{Observations and data analysis}
\label{sec:obs}

\subsection{High resolution spectroscopy}

\subsubsection{HARPS-N spectra}

After the studies of the system by \cite{desidera2014} and \cite{damasso2015}, spectroscopic monitoring of XO-2S and XO-2N was continued in the framework of the General Architecture of Planetary Systems (GAPS) program \citep{covino2013} to properly characterize the long-term trends observed for both components. Observations were performed with the High Accuracy Radial velocity Planet Searcher for the Northern emisphere (HARPS-N) spectrograph at the Telescopio Nazionale Galileo \citep[TNG, ][]{Cosentino2012}, with a typical integration time of 900 s for each component. For XO-2N, we collected 39 spectra between October 6, 2014, and May 10, 2023. For XO-2S, we gathered 106 spectra from September 10, 2014, to April 5, 2023. 

Radial velocities (RVs) were obtained with the HARPS-N pipeline using the online tool YABI\footnote{\url{http://ia2-harps.oats.inaf.it:8000/login/?next=/}}, using a K5 mask for the cross-correlation function (CCF). The pipeline provides additional indicators used in our analysis such as the Bisector Inverse Span (BIS) and the full width at half maximum (FWHM) of the CCF. We also exploited the YABI tool, with the procedure based on \cite{Lovis2011}, to measure the S Index, calibrated onto the Mount Wilson scale, and the chromospheric activity index \logrhk. In addition, we derived the H$\alpha$ and NaI activity indicators, both extracted from the spectra using the Python tool \texttt{ACTIN 2}\footnote{\url{https://actin2.readthedocs.io/en/latest/index.html}} \citep{daSilva2018, dasilva2021}. 

The basic properties of the RV time series are summarized in Table \ref{datasets}, while the individual HARPS-N RV time series are available at the CDS.
For XO-2N, we found a few outliers that we had to remove. In particular, we removed the spectrum taken on 18 October 2016 because it was contaminated by the Sun, the one taken on 06 January 2013 because the integration time was 1 s, and finally, the one taken on 27 December 2012 because of a technical issue.


\subsubsection{Archive and literature data}

As for the RVs, there are data available for XO-2N taken with the High Resolution Echelle Spectrometer (HIRES) by \cite{Butler2017}, the High Dispersion Spectrograph (HDS) by \cite{narita2011}, and with the Spectrographe pour l’Observation des Phénomènes des Intérieurs stellaires et des Exoplanètes (SOPHIE)\footnote{\url{http://atlas.obs-hp.fr/sophie/}}. Additionally, there are the original data taken by \cite{burke2007} with the High-Resolution Spectrograph (HRS) on the McDonald Observatory 11-m Hobby-Eberly Telescope (HET) for both stars but we did not include them because of their low quality compared to the other sets. After a few tests, we eliminated the SOPHIE data from our analysis because of their lower precision (the resulting RV jitter was $\sim$ 38 m/s, while for the other datasets it was $< 10$ m/s). The HIRES data include values of the S$_{MW}$ activity index, while the HDS data do not. The authors state that their S-index values are already calibrated in the  Mount Wilson scale, so they can be directly compared with the HARPS-N ones.  
No error bars are reported for the HIRES S$_{MW}$. Therefore, we took those points that were grouped within a few days and calculated the standard deviation. After doing so for two clusters of points, we made the average of the sigmas and used the resulting value as the error bar for the whole set. This implies that the obtained error bars are likely an overestimate of the instrumental error and might be affected by a rotational modulation. 
Of the HDS RV data set, we only kept those points that were not affected by the Rossiter-McLaughlin effect (\citealt{rossiter1924} and \citealt{mclaughlin1924}). \\
The properties of the datasets are shown in Table \ref{datasets}.


{\begin{table*}
\centering
\caption{\label{datasets}Properties of the datasets}
\begin{tabular}{ c c c c c c c c }
\hline
\\
Target & \multicolumn{2}{c }{HIRES} & \multicolumn{2}{c }{HDS} & \multicolumn{2}{c }{HARPS-N} & $\Delta t$ [yr] \\
\\
\hline
\\
 & N° points & $\langle \sigma_{RV} \rangle$ [m/s] & N° points & $\langle \sigma_{RV} \rangle$ [m/s] & N° points & $\langle \sigma_{RV} \rangle$ [m/s] & \\
\\
\hline
\\
XO-2N & 11 & 2.3 & 9 & 3.3 & 65 & 3.3 & 15.6 \\
\\
\hline
\\
XO-2S & - & - & - & - & 169 & 2.8 & 10.0 \\
\\
\hline
\\
\end{tabular} \\
Number of data points and corresponding average RV error bar for each target and instrument. The last column displays the total time span covered by the observations.
\end{table*}}



\subsection{Photometric data from TESS}
Our targets have been observed with the Transiting Exoplanet Survey Satellite \citep[TESS,][]{ricker2015} in sectors 20, 47, and 60 in a 2-minute cadence. In the TESS catalog, the two stars are labeled as TIC 356473034 (XO-2N) and TIC 356473029 (XO-2S). The light curves are obtained with the pipeline developed by the Science Processing Operations Centre \citep[SPOC,][]{jenkins2016}. The Simple Aperture Photometry (SAP) light curves have been downloaded with the \texttt{lightkurve} Python package \citep{lightkurve2018}. We decided to exploit the XO-2S data to check whether we can reveal the transits of XO-2S b and c, and the XO-2N data to try to constrain the rotation period of this component. The main problem is that the two stars are partially blended, and therefore the analysis requires thorough care. In particular, the transits of the planets of XO-2S might be diluted by the light of XO-2N.

\subsection{Data analysis tools}
\subsubsection{Periodograms}
We extracted periodograms using the \texttt{astropy} Python package. For XO-2N we first extracted the Generalized Lomb-Scargle periodogram \citep[GLS,][]{gls} of the HARPS-N data alone to estimate the offset and subtract it before doing the same with the complete RV time series. We did not follow the same procedure for the HIRES and HDS sets because there are too few points and such a process could lead to the determination of an inaccurate offset. This issue is not present for XO-2S as only HARPS-N data are available for this target. In all cases, the reported False Alarm Probabilities (FAPs) have been calculated with the bootstrap method.

\subsubsection{Orbital fit}
For the data analysis in this work, we used the 9.1 version of the Python tool \texttt{PyORBIT}\footnote{\url{https://pyorbit.readthedocs.io/en/latest/index.html}} \citep{Malavolta2016, Malavolta2018}, a package for the Markov Chain Monte Carlo (MCMC) modeling of RVs, activity indicators, and light curves. In particular, it is based on the combination of the optimization algorithm \texttt{PyDE}\footnote{\url{https://github.com/hpparvi/PyDE}} \citep{storn1997} and the affine invariant MCMC sampler \texttt{emcee}\footnote{\url{https://emcee.readthedocs.io/en/stable/}} \citep{foreman2013}. For the analysis of stellar activity data, we resorted to Gaussian Process (GP) regression with a quasi-periodic kernel as in the last years this technique has proven to be quite effective for this purpose \citep[e.g.,][]{Grunblatt2015}. In this context, the hyperparameter $h$ can be seen as the amplitude of the activity contribution to the RV signal, while $\theta$ is the period. In all cases, we ran the algorithm for 200000 steps with a burn-in parameter of 50000. For both RVs and activity indices, we introduced separate offset and jitter terms for HIRES, HDS, and HARPS-N data, respectively. We performed several fits for each target, selecting the best ones based on the Bayesian Information Criterion (BIC) value \citep{schwarz1978}. In addition, we also considered the Gelman-Rubin diagnostic \citep{gelman1992} to check for convergence of the parameters. In all cases, the reference time is BJD $= 2458842$. \\


\subsubsection{Light curve analysis}\label{lightcurveanalysis}
\begin{figure*}[t]
   \centering
   \includegraphics[width = 180 mm]{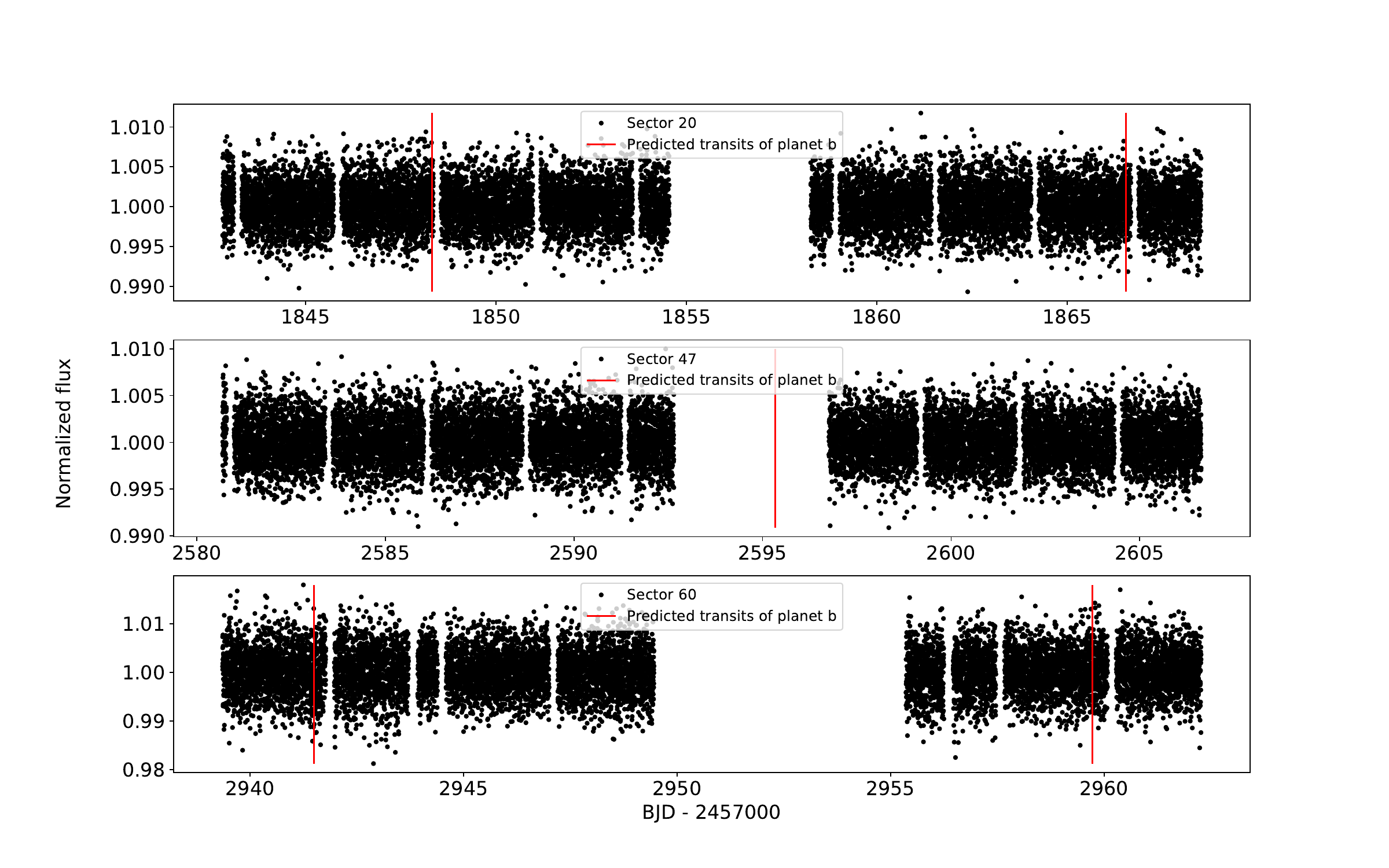}
      \caption{TESS light curves of XO-2S after removing outliers above 4 sigmas, normalizing, correcting for dilution, and removing the transits of XO-2Nb. The red bars indicate the predicted transits for planet b from RV results, as explained in Sect. \ref{transits}.
              }
         \label{fig:xo2srawlc}
   \end{figure*}

Before analyzing the TESS light curves with \texttt{PyORBIT} to search the signals corresponding to the planets known from RVs, we made some preparatory steps. First, we downloaded the SAP light curves, removing NaNs and outliers above 4 sigmas. As previously mentioned, the two components of the binary system are blended and, therefore, we estimated the dilution factor for XO-2S as described in Sect. 2.2.2 of \cite{mantovan2022}, obtaining a value of $0.376 \pm 0.003$ used to further correct our data and remove the contribution of the N component. Finally, we masked and removed the transits of XO-2N b. The light curves are shown in Fig. \ref{fig:xo2srawlc}.

\section{Stellar parameters}\label{sec:starparam}
The two stars of this binary system are both G9V main sequence stars at a distance of  $\sim 150$ pc. The two components have a separation of $\sim 31$ arcsec, implying a physical distance of $\sim 4600$ au. They have been thoroughly analyzed in many papers (\citealt{desidera2014}, \citealt{damasso2015}, \citealt{biazzo2015}, \citealt{teske2015}, and \citealt{zellem2015}) and the recent results from Gaia DR3 yield a trigonometric distance basically identical to the previously adopted one.
Therefore we did not perform any additional analysis to retrieve their physical parameters. In this work, we adopted the values reported in \cite{damasso2015} for our analysis. In particular, the masses and the radii are taken from their Table 2 (for XO-2N, we used the values marked with number 2 in the notes), while the other parameters are taken from the differential analysis results in Table 3 as this method has proven to be very effective in this case (see the original paper for an in-depth discussion). Such values are shown in Table \ref{tab:starparam}.

\begin{table}
\centering
\caption{\label{tab:starparam} Physical stellar parameters as derived in \cite{damasso2015}}
\begin{tabular}{ c c c }
\hline
\\
 Parameter & XO-2N & XO-2S \\
\\
\hline
\\
Radius [\rsun] & $0.998 \pm 0.033$ & $1.02 \pm 0.09$ \\
\\
Mass [\msun] & $0.96 \pm 0.05$ & $0.98 \pm 0.05$ \\
\\
\teff [K] & $5290 \pm 18$ & $5325 \pm 37$ \\
 \\
\logg [cgs] & $4.43 \pm 0.10$ & $4.420 \pm 0.094$ \\
\\
\feh [dex] & $0.37 \pm 0.07$ & $0.32 \pm 0.08$ \\
\\
\hline
\end{tabular}\\
\vspace{0.2cm}
\end{table}


\section{XO-2S: Evidence for a new Jupiter-analog planet}
\label{sec:xo2s}

\subsection{Previous works}
The two planets in the XO-2S system were announced by \cite{desidera2014}. Planet b is a warm Saturn-like planet with $P \sim 18$ d and $m\sin i \sim 0.26$ \mjup, while planet c is a temperate Jovian with $P \sim 126$ d and $m\sin i \sim 1.37$ M$_J$. 
A long-term RV trend was also noticed. In that work and also in \cite{damasso2015} it has been noted that activity does not seem to contribute to the RV signal. Nevertheless, since in the last 9 years we have expanded our data set with 106 more spectra, we considered its inclusion in the modeling of the time series.

\subsection{Periodogram analysis of XO-2S}
\begin{figure*}[t]
   \centering
   \includegraphics[width = 180 mm]{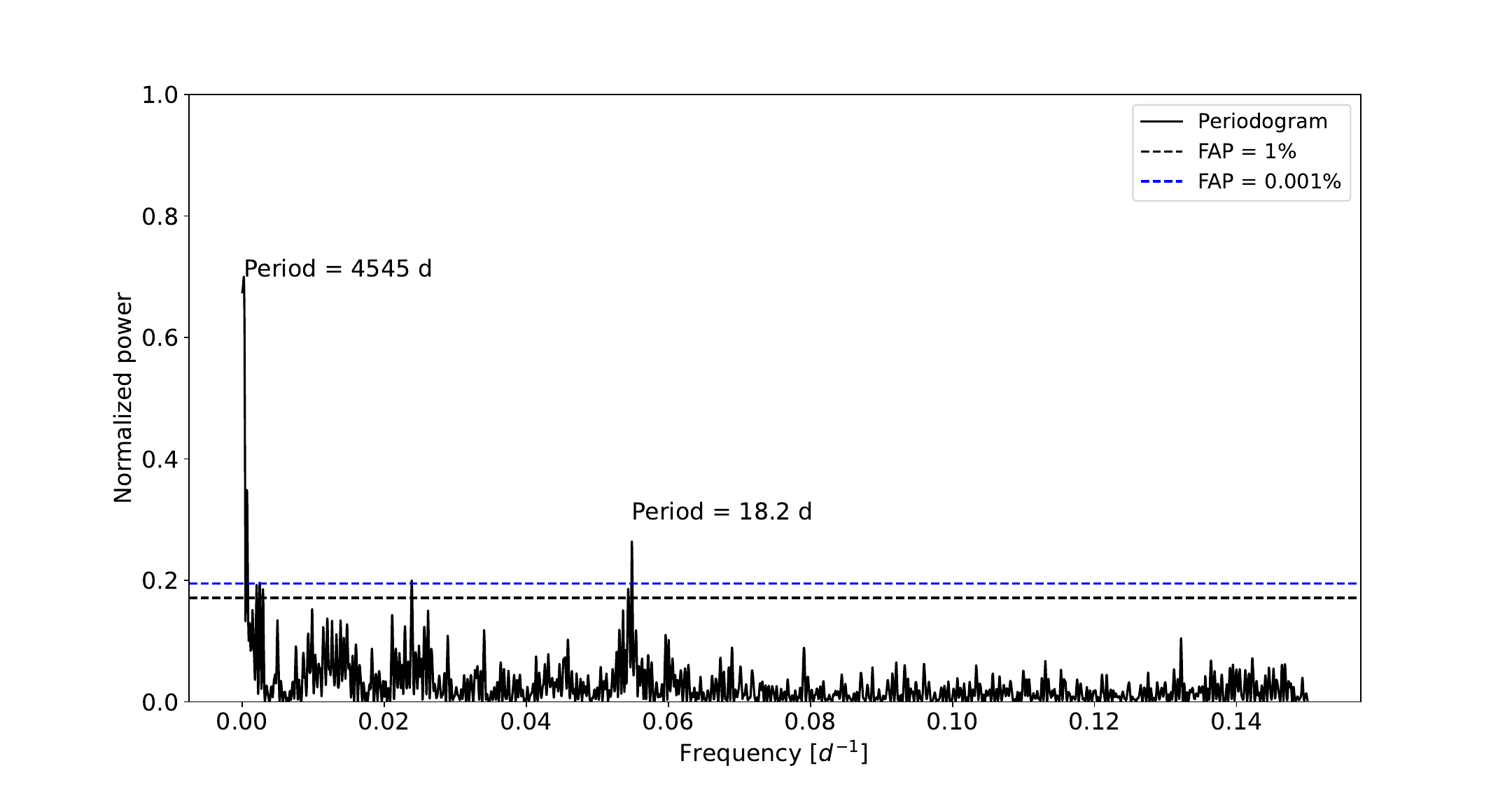}
      \caption{GLS of the RV residuals of XO-2S after removing the signal corresponding to planet c. The peak at 18.2 d corresponds to planet b.
              }
         \label{fig:xo2sglsrvres}
   \end{figure*}

The GLS computed from the RVs of XO-2S reveals a clear peak corresponding to planet c. The residuals show a significant peak at $\sim 4500$ d, as shown in Fig. \ref{fig:xo2sglsrvres}, revealing the presence of a potential additional planet. The other significant peak at larger frequencies corresponds to the signal of planet b.
\noindent The periodogram of the activity index S$_{MW}$ time series shows a peak at 3846 d with FAP $\sim 1\%$. In spite of that, we included this activity index in our fit to evaluate the contribution of stellar activity to the RV signal. We computed the GLS of two more activity indicators, the BIS and the FWHM of the CCF, but did not retrieve any information worth mentioning.

\subsection{Orbital fit of XO-2S}
\label{sec:fit_xo2s}

After the periodogram analysis, we proceeded with the full orbital analysis with \texttt{PyORBIT}. To fully explore the parameter space, we performed the following orbital fits:

\begin{enumerate}[i]
    \item Two planets without activity: this includes the two known planets and has been used as a reference (Model 1, M1).

    \item Three planets without activity: since there is a long-period signal, we fit it with a Keplerian (Model 2, M2).

    \item Two planets with quasi-periodic GP: this has been done to check whether the long-term signal might be due to an activity cycle rather than an additional planet (Model 3, M3).

    \item Three planets with quasi-periodic GP: to explore the possibility that there is an additional planet but the star also experiences an activity cycle that contributes to the RV signal (Model 4, M4).

\end{enumerate}

\begin{figure*}[t]
   \centering
   \includegraphics[width = 180 mm]{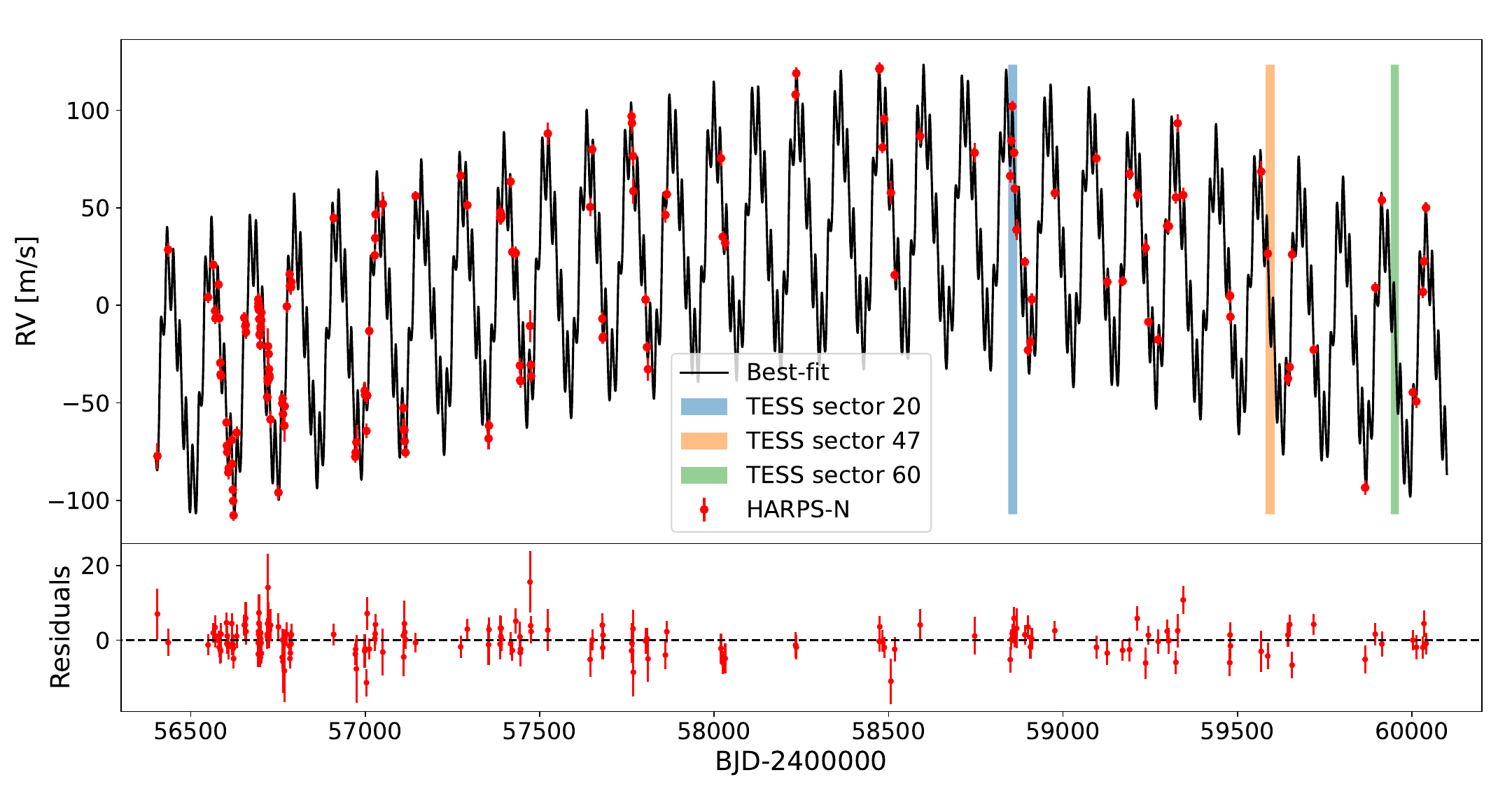}
      \caption{Best-fit model (Model 2) of XO-2S.
              }
         \label{fig:xo2sbestfit}
   \end{figure*}

\begin{figure*}[t]
   \centering
   \includegraphics[width = 180 mm]{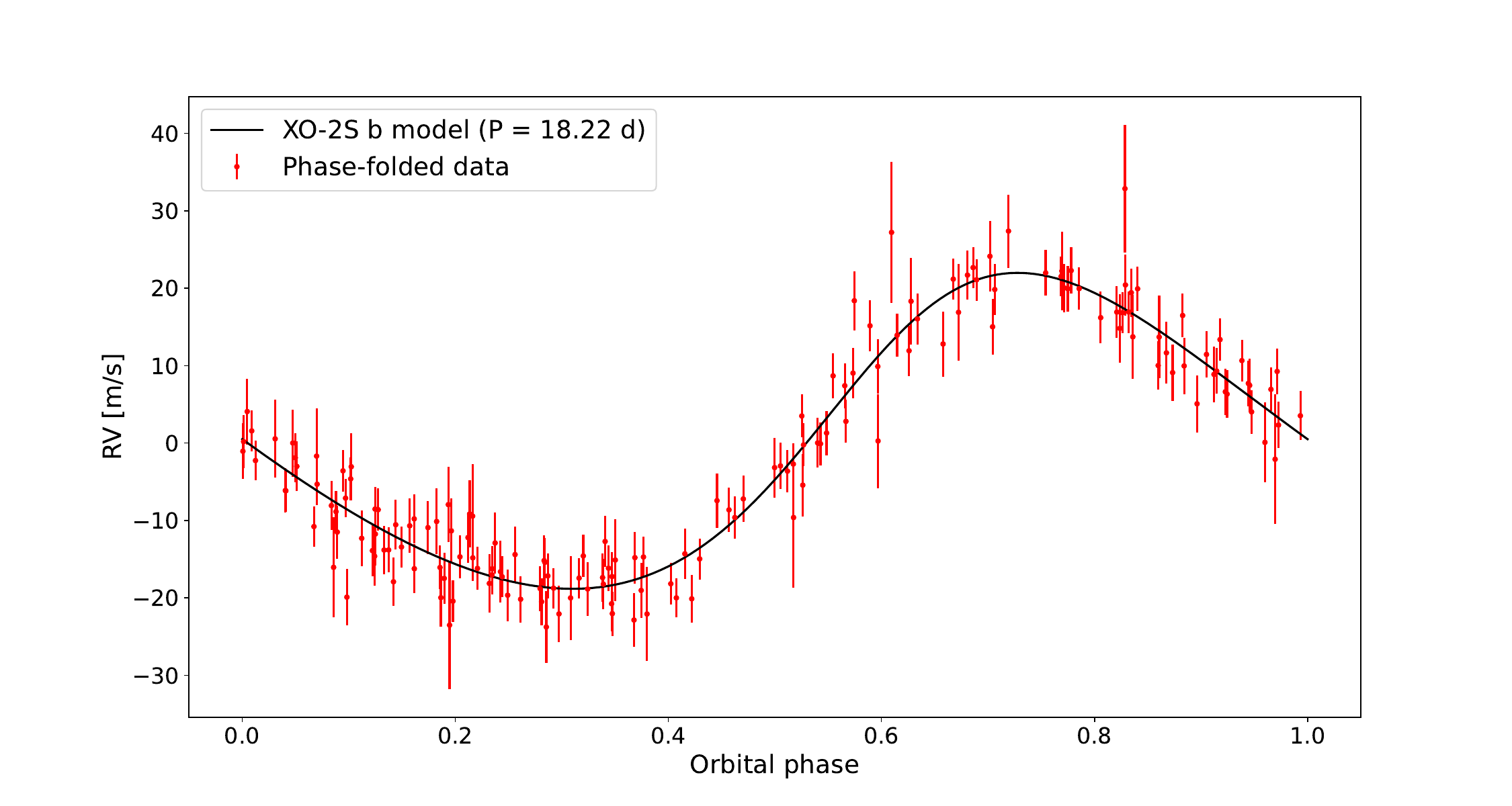}
      \caption{Best-fit Keplerian model for XO-2S b (phase-folded), after removing the RV signal of the other two planets.
              }
         \label{fig:xo2skepb}
   \end{figure*}

\begin{figure*}[t]
   \centering
   \includegraphics[width = 180 mm]{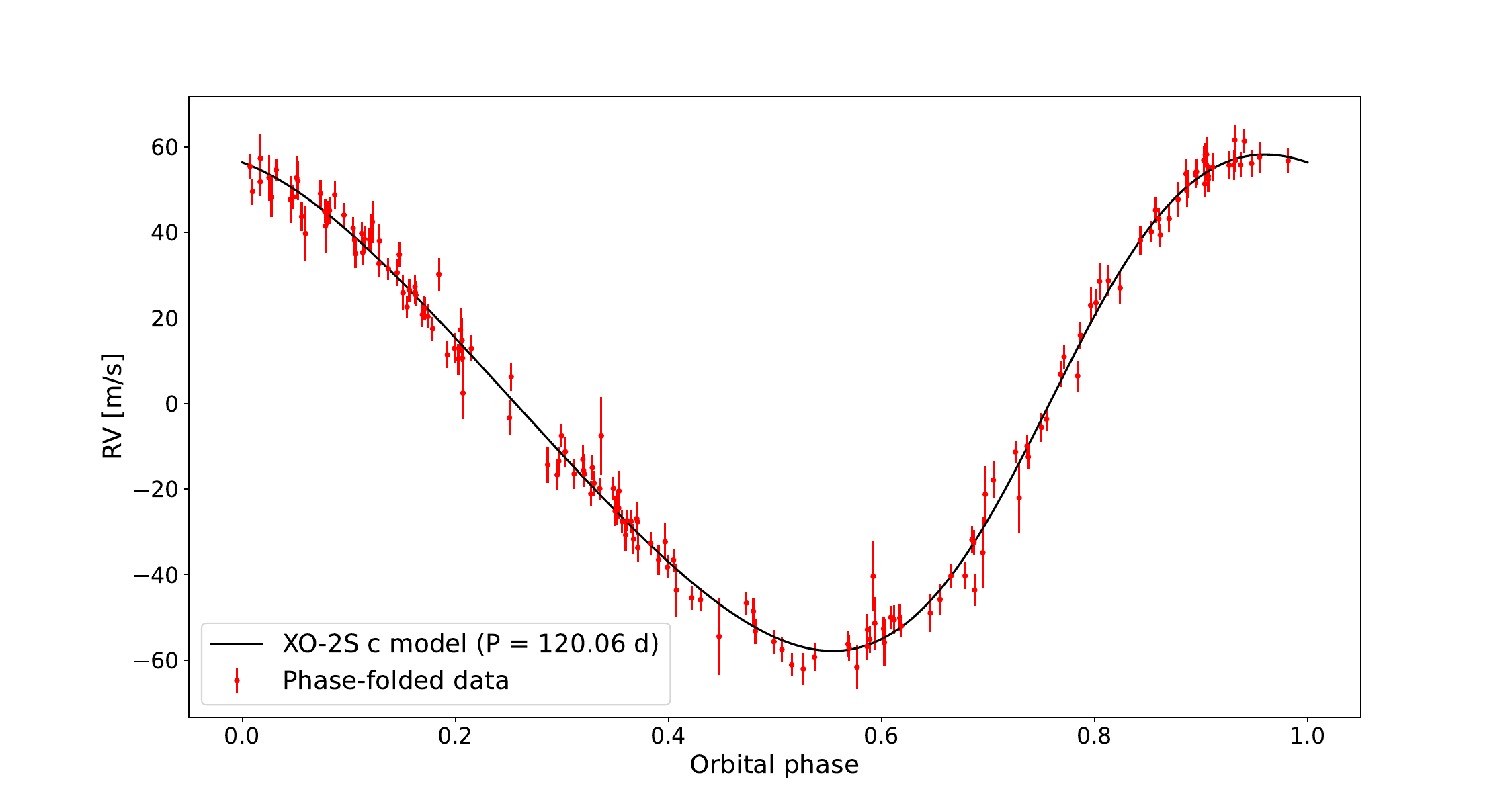}
      \caption{Best-fit Keplerian model for XO-2S c (phase-folded), after removing the RV signal of the other two planets.
              }
         \label{fig:xo2skepc}
   \end{figure*}

\begin{figure*}[t]
   \centering
   \includegraphics[width = 180 mm]{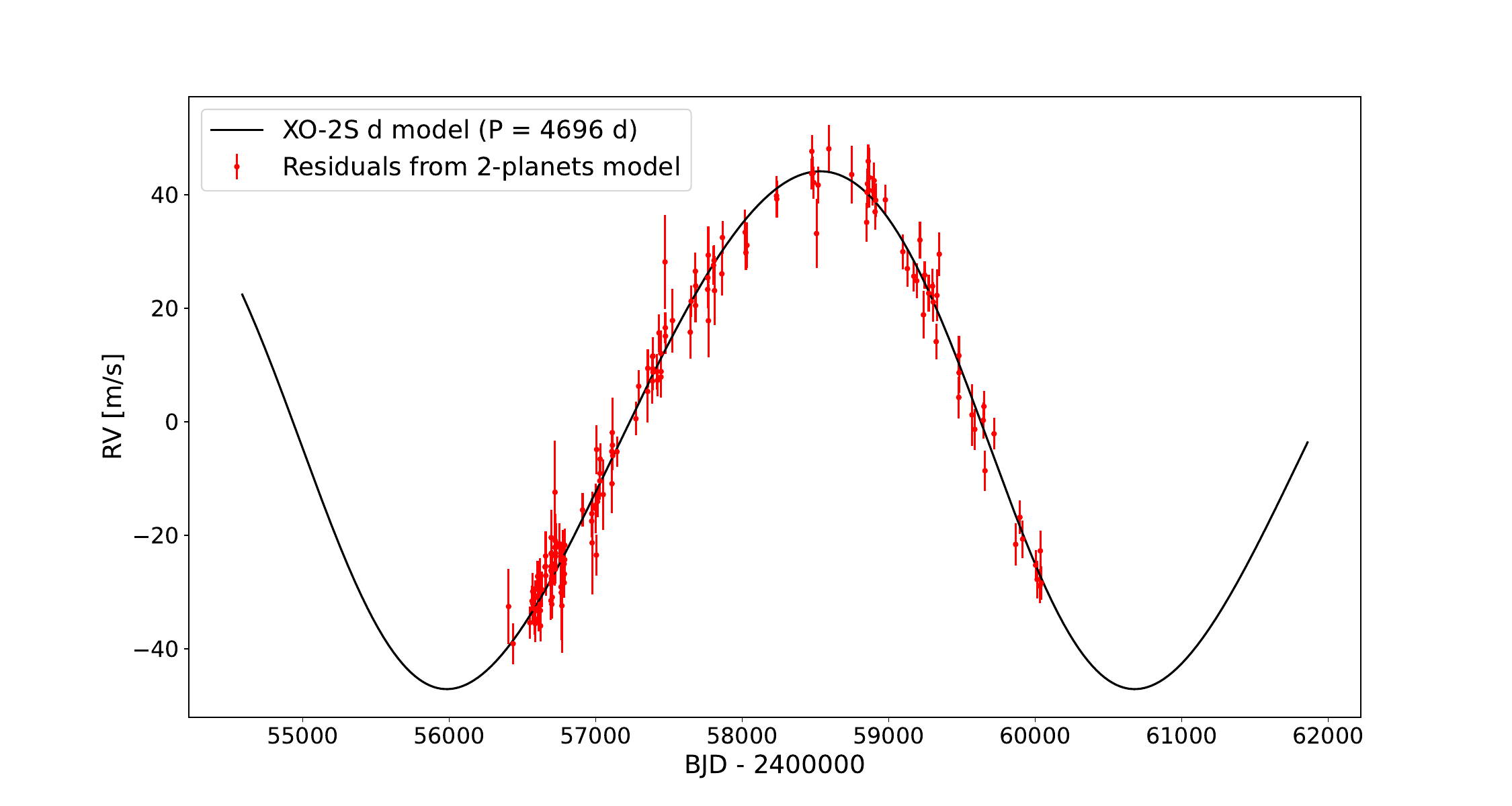}
      \caption{Best-fit Keplerian model for XO-2S d, after removing the RV signal of the other two planets.
              }
         \label{fig:xo2skepd}
   \end{figure*}

\begin{figure*}[t]
   \centering
   \includegraphics[width = 180 mm]{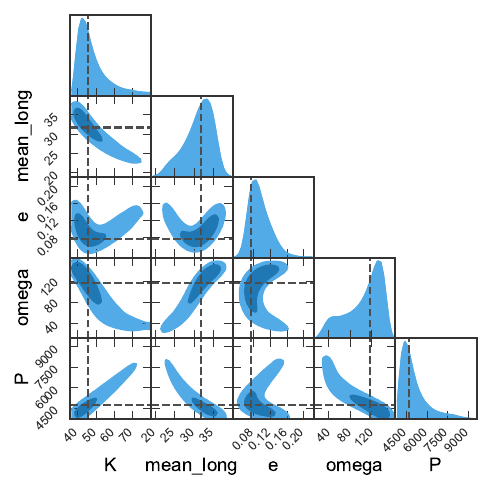}
      \caption{Corner plot of the orbital parameters of the new candidate XO-2S d obtained from our analysis.
              }
         \label{fig:cornerplot}
   \end{figure*}

We obtain BIC(M1) $= 663.0$, BIC(M2) $= 22.5$, BIC(M3) $= 44.1$, and BIC(M4) $= 53.1$. We consider the model selection criterium by \cite{kass1995} according to which a model is strongly favored over another when $\Delta \text{BIC} > 10$. In our cases, we have $\Delta$BIC = 21.6, 30.6, and 640.5 in favor of M2 compared to M3, M4, and M1, respectively, and therefore conclude that there is strong evidence that Model 2 is the most convincing representation of the actual RV data. The best-fit model (Model 2) is shown in Fig. \ref{fig:xo2sbestfit}, while Figures \ref{fig:xo2skepb}, \ref{fig:xo2skepc}, and \ref{fig:xo2skepd} show the Keplerian models of planet b, c, and d, respectively. The planetary parameters are shown in Table \ref{tabparam}. As we can see in Fig. \ref{fig:xo2skepd}, unluckily our time series does not sample any minima of the signal corresponding to the third object. This implies a correlation between $K_d$ and $P_d$, as shown in the corner plot in Fig. \ref{fig:cornerplot}. Therefore, our results on $m_d \sin i_d$ are likely underestimated and should be considered as a minimum value. For this reason, we performed an additional fit including in the model two Keplerians and a quadratic trend (Model 5, M5) to see whether we could fit the data with a simple polynomial term instead of a full Keplerian curve. With this model, we obtained BIC(M5) $= 171.5$ and thus discarded it as unfavored. In addition, in such a model, the error bars of the orbital parameters of the known planets are 50-60\% larger and the jitter term is more than twice (4.8 vs 2.1 m/s) compared to the 3-planets case. \\
In summary, we confirm the two planetary companions XO-2S b and c originally discovered by \cite{desidera2014}, with minor differences in the orbital parameters but much smaller error bars (ranging from 3\% to 70\% of the older value for $P_b$ and $m_b \sin i_b$, respectively), and we identified a new planetary companion, to be named XO-2S d, in a Jupiter-like orbit ($a = 5.46$ au, $e = 0.091$) with a minimum mass of 3.71 \mjup. In the GLS of the RV residuals, the highest peak is at 30 d with FAP = 7.4\%, thus we have no evidence for additional planets in the system with the current data set.

\subsection{Search for transits of  XO-2S b and c in TESS data}
\label{transits}
So far, in the literature, there have been no dedicated transit searches of the two inner planets of XO-2S. For this reason, we looked for potential transit signals in the TESS light curves. Since planet b is quite close to its star ($a \sim 0.13$ au), the search for its transit is promising. Planet c is 3.5 times more distant; therefore, only a smaller range of inclinations would allow it to transit. We note here that, assuming a radius equal to that of Jupiter and $i = 90^{\circ}$, the transit durations are around 6 and 11 hours for the two planets.

Since our RVs results are robust and the error bars on the orbital parameters are pretty low, we used them to predict where the potential transits could occur within the TESS sectors (see, e.g., \citealt{kane2007}). The uncertainty on the transit times is calculated as 

\begin{equation}\label{eqn:errtransit}
    \sigma = \sqrt{\sigma_{T_0}^2 + (\sigma_P \cdot n)^2}
\end{equation}

\noindent where $\sigma_{T_0}$ is the uncertainty on the time of the first transit, $\sigma_P$ the error on the orbital period, and $n$ an integer indicating the number of periods passed between the reference ephemerids and the time of TESS data. Rigorously speaking, one should include the light-time effect produced by the motion of XO-2N around the barycenter of the visual binary when estimating the difference in the times of mid-transits. Nevertheless, an order of magnitude estimate shows that the amplitude of the light-time effect is on the order of 1 minute over a time span of ten years, that is, it is negligible compared to the variation associated with the uncertainty of the orbital period of the planet. The values for $\sigma_{T_0}$ and $\sigma_P$ are reported in Table \ref{tabparam}, and we used BJD $= 2458842$ as reference time since it corresponds to the beginning of the TESS observations in Sector 20. In particular, we find for planet b a total of 5 potential transits at BJD $= 2458848.32$, $2458866.54$, $2459595.33$, $2459941.50$,  and  $2459959.72$, with an error of 0.4 d for Sector 60. However, the third predicted transit falls inside the TESS instrumental gap in Sector 47 (as shown in Fig. \ref{fig:xo2srawlc}) and so it must not be considered. In Eq. \ref{eqn:errtransit}, we used the error on the time of periastron passage instead of $\sigma_{T_0}$ because we have no previous known transit. Unfortunately, none of the predicted transits for planet c fall in the range covered by the TESS data, even considering the error bar of 1 d. For the sake of completeness, we also considered the possibility that the new candidate is transiting. We calculated the predicted times of transits and the associated uncertainties using the full posterior distribution of its orbital parameters. We found that, considering the uncertainty, the predicted time of transit is at least 500 d before the beginning of Sector 20, and the next one falls almost 2400 d after the end of Sector 60. Therefore, we carried on with the analysis searching for one potentially transiting planet. \\
As a first step, we used the Python package \texttt{transitleastsquares} \citep{hippke2019} to compute the Transit Least Squares (TLS) of the TESS light curves to detect any potential planetary signal. Unfortunately, we found no significant signal but proceeded with a full analysis to make sure we did not leave anything behind and, for example, a very grazing transit could perhaps remain undetected with the TLS approach. Before proceeding with the \texttt{PyORBIT} analysis, we estimated the limb-darkening coefficients (quadratic law) using the Python code \texttt{PyLDTk} \citep{Parviainen2015}, based on the spectrum library by \cite{Husser2013}. Using the effective temperature, $\log g$, and metallicity reported in Table \ref{tab:starparam} for XO-2S, we obtained the values $u_1 = 0.4690 \pm 0.0066$ and $u_2 = 0.1119 \pm 0.0198$, and we set them as priors in our light curve analysis.
In addition, we set the following priors: the period of the planet as obtained from the RV analysis, the transit time given by the first predicted transit, and the stellar density ($0.925 \pm 0.25\ \rho_{\odot}$) using the mass and radius in Table \ref{tab:starparam}. We also fixed the eccentricity and $\omega$ values since these are not retrieved from the light curves.\\
Unfortunately, our results indicate that a transit of XO-2Sb is very unlikely to be present in the TESS data. In particular, we obtain an impact parameter $b = 1.00_{-0.52}^{+0.31}$ and a planetary radius $R_p = 0.41_{-0.32}^{+2.9}$ R$_J$. The nominal values are clearly nonphysical (the radius implies an Earth-like density for a gas giant), and the extremely large error bars are due to posterior distributions that are either almost flat (impact parameter) or peaked at very low values and then flat up until very large ones (planetary radius). Therefore, we must conclude that, at present, the evidence is that XO-2S b is non-transiting. As said before, we cannot say anything about XO-2S c and d but if their orbits are coplanar to that of planet b (or very close to it), then they are also very unlikely to transit. \\
As a final step, we checked the sensitivity of TESS data to the presence of planet b. In practice, we injected fake transits at the times predicted by RV results with the Python package \texttt{batman} \citep{kreidberg2015} and tried to recover them with the TLS. We used $R_p/R_s = 0.12$ (corresponding to roughly Jupiter's radius) and repeated the procedure with decreasing radii until the transit was no longer recovered significantly, that is when SED $\lesssim 10$. For this analysis, we fixed $i = 90°$, while also noticing that the transit condition for this planet is $88^{\circ} \leq i \leq 92^{\circ}$ based on its RV-derived orbital parameters. We note that, at the first step, we obtain SED $= 24.9$. We continued the analysis down to $R_p/R_s = 0.04$ (roughly Neptune's radius), in which case we found SED $= 9.5$. These results indicate that, with the current TESS data, we are potentially able to detect a transiting planet slightly larger than Neptune. Since the minimum mass of XO-2S b is 0.26 \mjup, we expect its radius to be $\sim 1$ R$_J$ and this further confirms that, if it was transiting, we would be able to see it.

\subsection{Dynamical stability of the system}
\label{sec:stability}

The long-term stability of the system has been tested with the chaos indicator MEGNO, the Mean Exponential Growth factor of Nearby Orbits (\citealt{MEGNO1} and \citealt{MEGNO2}) closely related to the maximum Lyapunov exponent. It is computed by solving the variational equations of the dynamical system to evaluate the relative divergence of orbits. The value of MEGNO in the case of regular or quasi–periodic orbits is very close to 2, while for chaotic motion it increases with time. The system of three planets for the nominal values of the masses and orbital parameters given in Table \ref{tabparam} is stable, according to MEGNO. We increased the planets' masses up to 10 times their nominal values, assuming the planets are coplanar (corresponding to an inclination of ($i \sim 6^{\circ}$), but the system remains stable. 

\begin{figure*}[t]
   \centering
   \includegraphics[width = 180 mm]{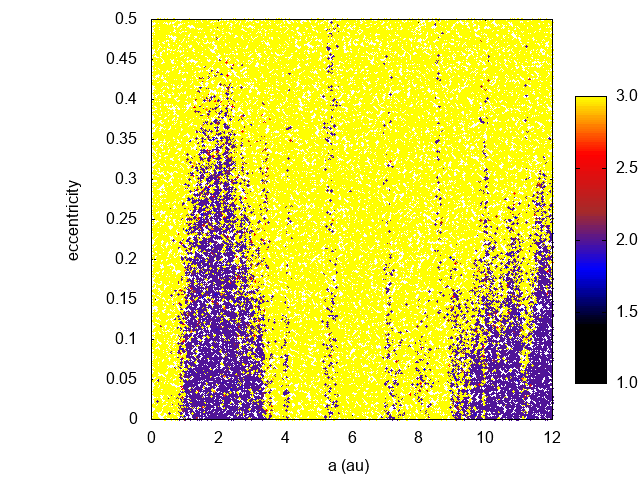}
      \caption{Stability results using the chaos indicator MEGNO (color bar on the right) for XO-2S, using the nominal values for the masses of the three planets.
              }
         \label{fig:xo2s_stability}
   \end{figure*}

\begin{figure*}[t]
   \centering
   \includegraphics[width = 180 mm]{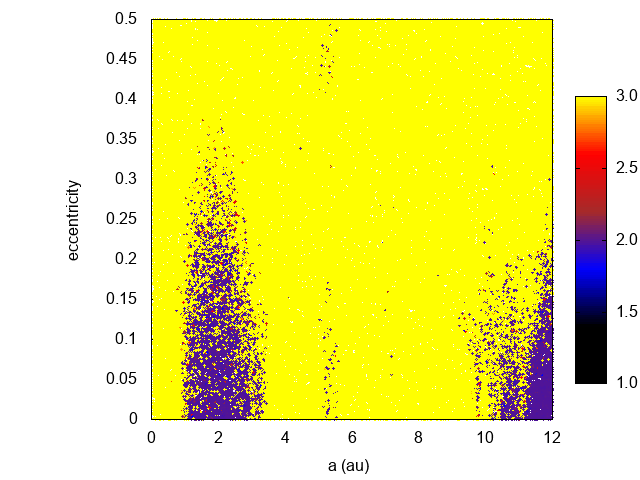}

      \caption{Stability results using the chaos indicator MEGNO (color bar on the right) for XO-2S, increasing the masses of the three planets by a factor of 3 with respect to the $m\sin i$.
              }
         \label{fig:xo2s_stability_10m}
   \end{figure*}

We also investigated the location of a potential additional planet on a stable orbit. We have assumed an Earth mass for this putative planet and randomly sampled its orbital elements. The semi-major axis is varied between 0.05 and 12 au while the eccentricity ranges from 0 to 0.5. Stable regions which could harbor an additional Earth-like planet are found mainly in the 1-3 au range, and beyond 9 au, as shown in Fig. \ref{fig:xo2s_stability}. There are also some stable orbits corresponding to Mean Motion Resonances.
If we increase the masses by a factor of 3 (Fig. \ref{fig:xo2s_stability_10m}), the prominent stable regions get smaller and the resonant orbits are no longer stable. The situation worsens if we increase the masses by a factor of 10, even though the main stable regions remain. Unfortunately, as we describe in Sect. \ref{astrometry_imaging}, astrometry does not help us to rule out the low inclinations ($i \sim 6^{\circ}$) that this hypothetical scenario implies. 

As a final case, we focused on the region at $a < 0.5$ au, increasing the spatial sampling. Despite a few isolated stable cases, we find that most of this region is unstable, especially between the two inner planets. These results have been obtained by fixing the masses to the $m\sin i$. 


\section{XO-2N: An activity cycle inducing a planetary-like signal}
\label{sec:xo2n}

\subsection{Previous works}
XO-2N b has been discovered by \cite{burke2007} with the transit method and an RV follow-up. It is a hot Jupiter with $P = 2.6$ d and $M_p \sim 0.6$ \mjup. \cite{damasso2015} noted the presence of a long-term trend in the RV time series and pointed out the possibility that it was caused by a magnetic cycle of the star but at the time they did not have enough data points to confirm this statement. Now, we have 39 more spectra gathered over the last 9 years and even more accurate ephemeris for planet b thanks to TESS data \citep{ivshina2022}, so we analyzed this system once again to determine the nature of the additional RV signal.

\subsection{Periodogram analysis of XO-2N}\label{xo2n_gls}

\begin{figure*}[t]
   \centering
   \includegraphics[width = 180 mm]{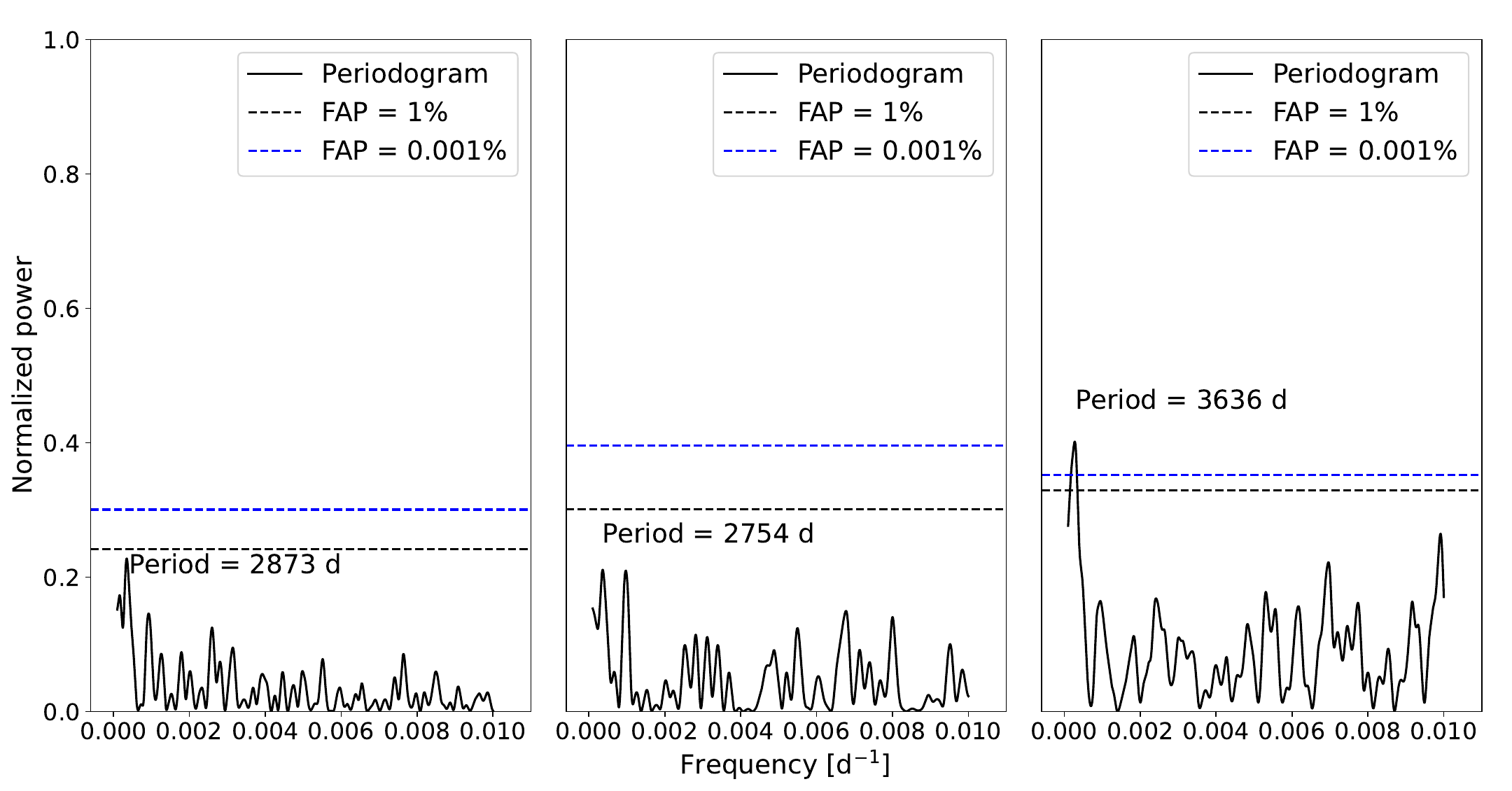}
      \caption{Periodograms of XO-2N. \textit{Left panel:} GLS of RV residuals of XO-2N obtained from a 1-planet model without activity. \textit{Central panel:} GLS of the BIS data of XO-2N. \textit{Right panel:} GLS of the S$_{MW}$ activity index of XO-2N. 
              }
         \label{fig:xo2nglsrvres}
   \end{figure*}

The periodogram of the RVs of XO-2N displays an extremely well-peaked signal at the period of planet b ($\sim 2.6$ d). 
The GLS of the RV residuals obtained by fitting a 1-planet model without any activity contribution displays a signal at a little less than 3000 days, even though it is not particularly strong (FAP $\sim 2\%$). A similar peak (2754 d) is found in the BIS data, even though its FAP is 25\% and therefore unreliable. This is probably due to the fact that the BIS data only come from the HARPS-N data set and therefore cover a much smaller time range. A similar peak ($P \sim 3600$ d) is found in the S$_{MW}$ time series (FAP $\sim 0.001\%$). These are shown in Fig. \ref{fig:xo2nglsrvres}. 
The combination of these three results suggests the presence of a common cycle with a period of about 9-10 yr. 


\subsection{Correlation between RVs and activity indicators}
\begin{figure*}[t]
   \centering
   \includegraphics[width = 180 mm]{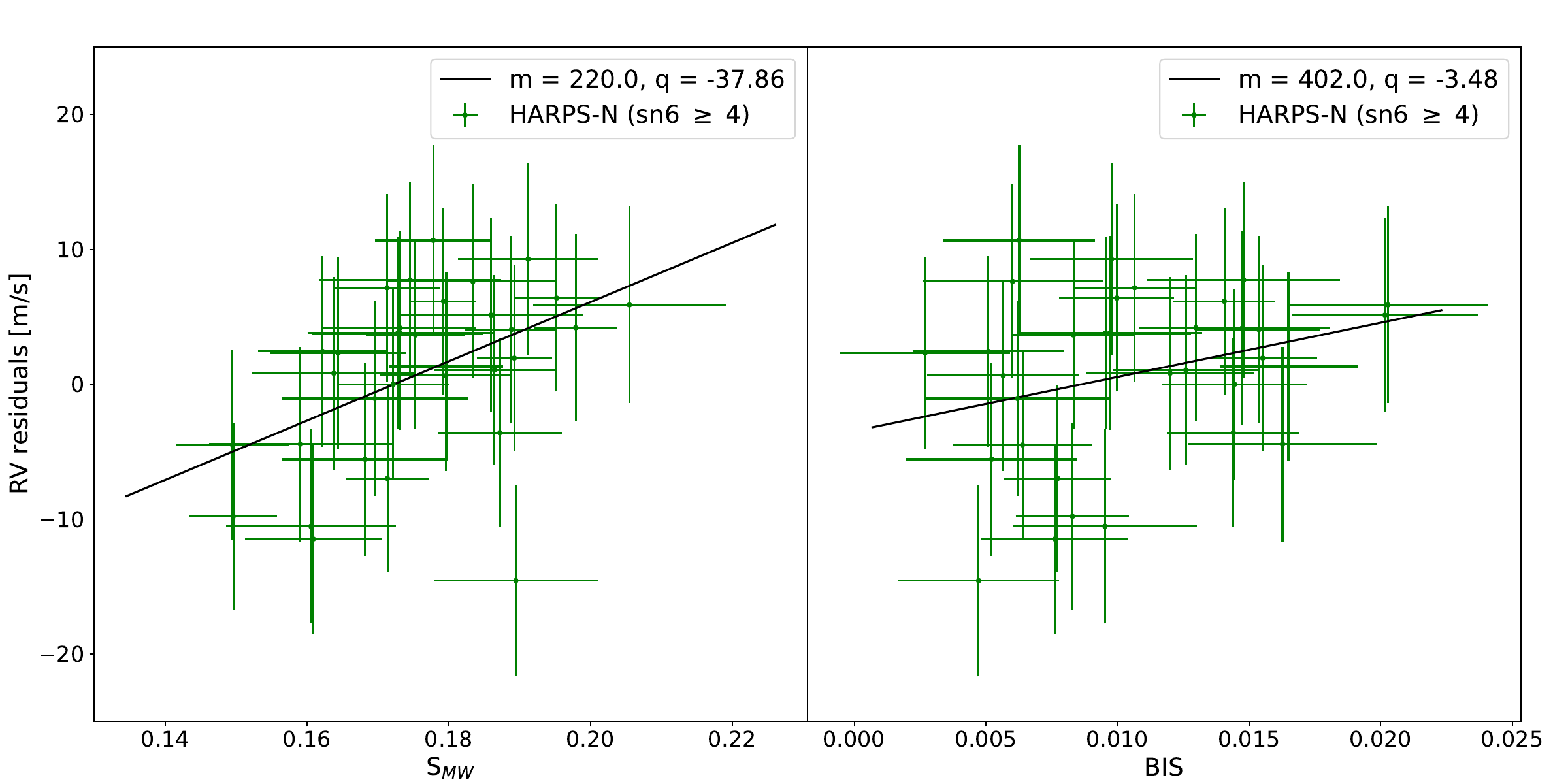}
      \caption{Correlations between RV residuals and activity indicators of XO-2N. \textit{Left panel:} Correlation between S$_{MW}$ and RV residuals. \textit{Right panel:} Same correlation but between the BIS indicator and RV residuals. 
              }
         \label{fig:smwbis}
   \end{figure*}

\cite{damasso2015} first noted that the long-term trend observed in the RVs of XO-2N could be due to an activity cycle. In particular, they showed that the residuals of the 1-planet model were strongly correlated with the activity index S$_{MW}$ (Spearman's rank $r = 0.68$, FAP $= 10^{-4}$). The result was obtained by selecting only the data points with a signal-to-noise ratio (S/N) $\geq 4$ in the 6th spectral order, which corresponds to the part of the spectrum where the Ca II H \& K lines are located. Following the same approach, we first fit the RV time series with a 1-planet model and then searched for the same correlation for different activity indices. Once again, we kept only the observations with S/N $\geq 4$. For this reason, we only used HARPS-N data for this analysis because we have more control over the observations and their technical information. The parameters obtained for planet b in the orbital fit are compatible with the ones found in previous works and are shown in Table \ref{tabparam}. \\
Fig. \ref{fig:smwbis} shows the correlations between the RVs and two activity indices, namely S$_{MW}$ and the BIS. As we can see, the correlation with the S-index is rather strong ($r = 0.51$ and $p = 0.0027$), even though nominally weaker than that found in \cite{damasso2015}. However, this is intrinsically more significant because it has been obtained using a larger data set over a much longer time range. The correlation with the BIS is weaker and less significant ($r = 0.31$ and $p = 0.08$), indicating that the activity signature in the RV is probably better mapped by the chromospheric features rather than photospheric ones (starspots). It is likely that the photospheric blue shift quenching is dominating the cyclic RV variations in this star, which is not particularly active, as it happens in the Sun \citep[cf. ][]{lanza2016}. Such a quenching is indirectly mapped by the chromospheric emission because it is produced by magnetic fields that are more diffuse and weaker than those in dark starspots. Those relatively weak fields manifest themselves as faculae and are associated with a chromospheric signature. 
We applied the same analysis using the H$\alpha$ and Na I indicators, obtaining $r = 0.35$ and $p = 0.05$ for both indices, indicating a weak correlation. Finally, we studied the correlation between RV residuals and the FWHM of the CCF. To do this, we had to remove the data points taken before March 2014 (14 spectra) because of a small defocus of the instrument, as explained in \cite{benatti2017}. With the remaining set, we obtain $r = 0.54$ and $p = 0.016$, once again proving a correlation between RVs and the star's activity.

\subsection{Rotation period}
As shown in the previous section, stellar activity is clearly affecting our RVs for XO-2N. 
While the periodograms presented in Sect. \ref{xo2n_gls} suggest that the most significant stellar variability is on a timescale of several years and then likely linked to an activity cycle, the investigation of the stellar rotation period is also highly relevant for a better understanding of the activity behavior of the star and to drive the most appropriate choices for the RV modeling (Sect. \ref{sec:cycle_GP}).

\cite{damasso2015} tried to determine the rotation period of this star using $\sim 800$ APACHE I-band data points, finding a value of $41.6 \pm 1.1$ d. This is consistent with the values expected from the $S_{MW}$ and color values, in particular, we obtain $39.6 \pm 0.6$ d using the expression by \cite{noyes1984} and $41.8 \pm 0.8$ using instead the formula by \cite{mamajek2008}. On the other hand, \cite{zellem2015} found low-amplitude brightness variations with periods of $29.9 \pm 0.16$ d in the 2013-2014 season and $27.34 \pm 0.21$ d in the 2014-2015 season. These values are incompatible with the previously mentioned results but, as the authors point out, still fall in the 29-44 range predicted by \cite{burke2007} based on radius and $v\sin i$. In our $S_{MW}$ data set, we do not see evidence for any rotational modulation because we are dominated by the long-period signal mentioned in Sect. \ref{xo2n_gls}, and our data are probably too sparse to reliably detect a rotational periodicity, considering the typical lifetime of active regions that does not exceed a couple of weeks in most cases, assuming the Sun as a template for such slowly rotating stars \citep[e.g.,][]{lanza2004}. 

To analyze the matter more in depth, we considered the PDCSAP TESS light curves. First of all, we removed the data points located inside the transits of XO-2N b using the \texttt{transitleastsquares} package. Then, we computed a GLS to search for periodicities that might correspond to the rotation period of the star. We repeated this procedure for each TESS sector separately and then for the whole light curve. In Sector 20, the peak corresponds to a 12.6-day signal that is too low to be a reliable estimate of the rotation period, even though it has FAP $\ll 0.001\%$. The same occurs in Sector 47, where the peak is at 10.1 days with FAP $\sim 0.4\%$. Sector 60 displays a 13.2-day significant peak that once again is unlikely to correspond to the star's rotation. The GLS extracted from the whole light curve is very noisy with high power between 10 and 15 days, and a nominal peak at 11.6 d so this does not further help to get information on the matter. However, all these signals probably have an instrumental origin. The conclusion is that the rotation period of XO-2N is at least dubious and probably longer than the duration of a TESS sector. 


\subsection{Orbital fit with multidimensional Gaussian Process analysis}
\label{sec:cycle_GP}

\begin{figure*}[t]
   \centering
   \includegraphics[width = 180 mm]{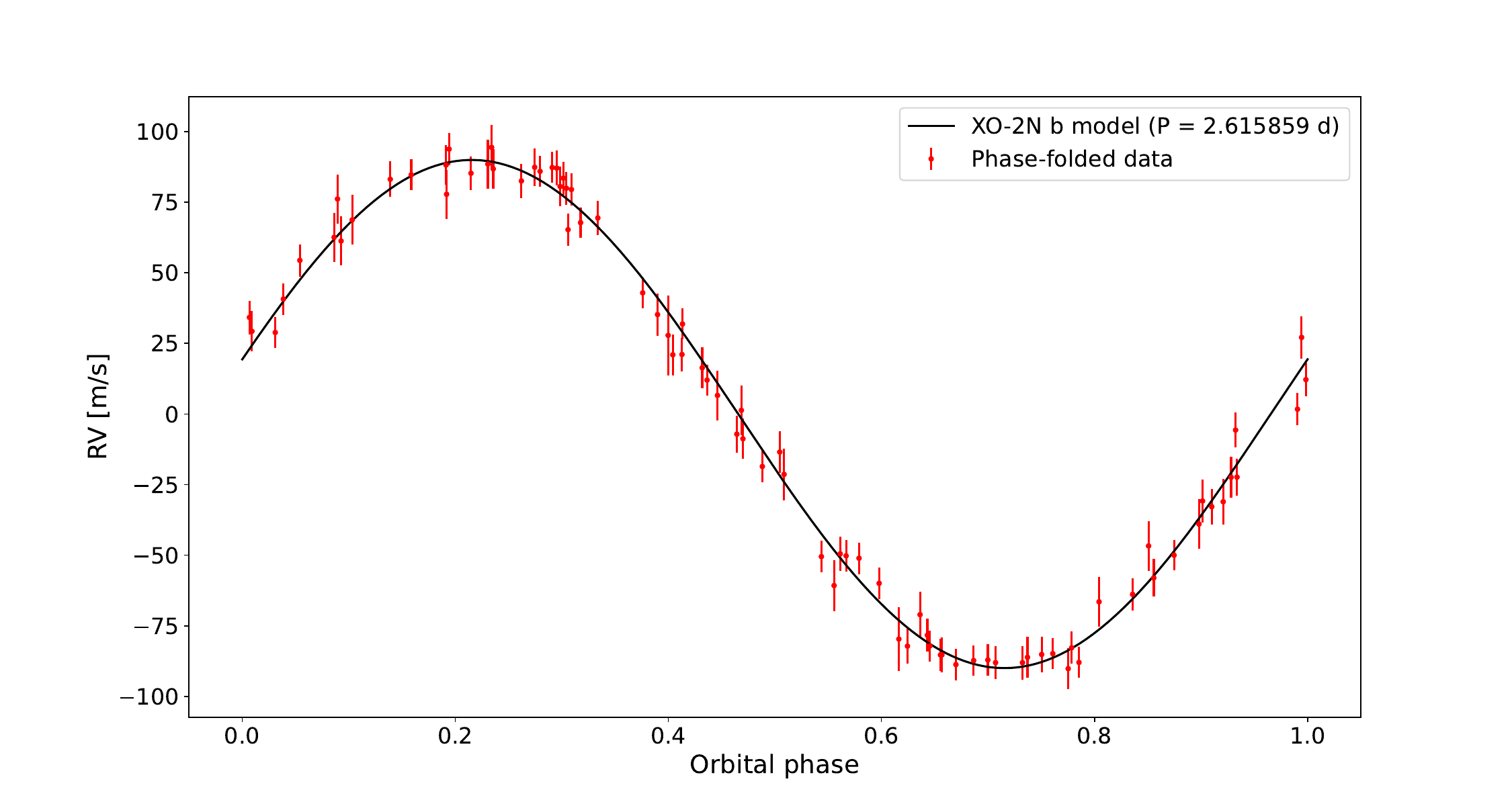}
      \caption{Best-fit Keplerian model for XO-2N b (phase-folded), after removing the multidimensional GP.
              }
         \label{fig:xo2nkepb}
   \end{figure*}

To further strengthen our evidence in favor of an activity cycle as the dominant source of the long-term RV variability, and to characterize it, we proceeded with the analysis using the multidimensional GP with a quasi-periodic kernel as described in \cite{rajpaul2015}. Shortly, this approach finds a function (the GP itself) shared between the RV, S$_{MW}$, and the BIS datasets, that is multiplied by suitable coefficients to reproduce the time series of these quantities. In practice, the system to solve is

\begin{equation}
\left\{ \begin{array}{l}
\text{RV} = V_c G(t) + V_r \dot{G}(t) + \text{planets}  \\
\\
S_{MW} = L_c G(t) \\
\\
\text{BIS} = B_c G(t) + B_r \dot{G}(t)
\end{array} \right.
\end{equation}

\noindent where $G(t)$ is the GP and $\dot{G}(t)$ its time derivative with $t$ being the time. The coefficients with subscript $r$ are rotational\footnote{This definition arises from the fact that in most cases, the GP regression analysis is performed for stars for which the dominant stellar contribution to RV variability is due to rotational modulation. Nevertheless, the same formalism can also be used for activity cycles.} terms, while those with subscript $c$ are related to the convective blueshift. See the original paper for more details. In this way, we can directly combine RV and two activity indicators to see if a common periodicity fits the data better than a simple Keplerian. After the first tests, we found that $V_r$ was completely undetermined (flat posterior distribution), and therefore we fixed it at zero, finding that its removal did not worsen our results. For planet b, we set a Gaussian prior on the period ($2.61585906\ \pm$ 2.5e-7 days) based on TESS light-curves \citep{ivshina2022} and used a circular orbit since, if a Keplerian is used, the eccentricity is very low ($\sim 0.006$) and $\omega$ is completely undetermined. As an alternative, we also tried to fit the RV data with two Keplerian terms. To be more rigorous, we derived the BIC for the model with the multidimensional GP using the RV part only. In this way, we can directly compare it with the 1-planet and the 2-planet models. We find BIC(1 planet) $= 22424$, BIC(1 planet + GP) $= 12338$, and BIC(2 planets) $= 19420$, so this indicates that there is strong statistical evidence in favor of the model including one planet and the multidimensional GP. Given these results, our interpretation is that there is no detectable second planet and the additional RV signal is due to an activity cycle with amplitude $V_c = 12.0_{-4.1}^{+4.6}$ m/s and period $P_{cycle} = 3378_{-189}^{+284}$ d. We also note that XO-2N is a star virtually identical to the Sun, except for its metallicity. Therefore, it is not surprising that it displays an activity cycle with a period (9-10 years) and amplitude of RV variations ($\sim 10$ m/s) similar to the Sun \citep{lanza2016}. The coefficients of the multidimensional GP and the period of the activity cycle are shown in Table \ref{tabgp}, while the parameters obtained from the RV analysis fit are shown in Table \ref{tabparam}. The best-fit model is shown in Fig. \ref{fig:xo2nkepb}.

 The best-fit model is shown in Fig. \ref{fig:xo2nkepb}.

\section{Discussion}
\label{sec:discussion} 

\begin{table*}
\centering
\caption{\label{tabparam} Planetary parameters obtained from the best-fit models.}
\begin{tabular}{ c c c c c }
\hline
\\
 Parameter & \multicolumn{3}{c }{XO-2S} & XO-2N \\
\\
\hline
\\
 & b & c & d & b \\
\\
\hline
\\
$m\sin i$ [M$_J$] & $0.26 \pm 0.01$ & $1.38 \pm 0.05$ & $3.71_{-0.51}^{+1.2}$ & $0.594 \pm 0.022$ \\
\\
P [d] & $18.220 \pm 0.001$ & $120.059 \pm 0.013$ & $4696_{-489}^{+1133}$ & $2.61585906 \pm 2.52\text{e}-7$ $^{(1)}$ \\
\\
a [a.u.] & $0.1347 \pm 0.0025$ & $0.4737_{-0.0088}^{+0.0085}$ & $5.46_{-0.40}^{+0.85}$  & $0.03666 \pm 0.00065$ \\
\\
e & $0.15 \pm 0.02$ & $0.149 \pm 0.006$ & $0.091_{-0.018}^{+0.028}$ & 0 (fixed) \\
 \\
$\omega$ [degrees] & $-59.1 \pm 7.0$ & $-88.6 \pm 2.8$ & $115_{-50}^{+22}$ & 90 (fixed) \\
\\
K [m/s] & $20.40 \pm 0.43$ & $58.0 \pm 0.4$ & $45.6_{-4.4}^{+10.0}$ & $90.0 \pm 1.0$ \\
\\
$T_0$ [BJD-2400000] & $58834.8 \pm 0.4$ & $58813.3 \pm 1.0$ & $59923_{-489}^{+131}$ & $58843.2176 \pm 0.0003$ $^{(1)}$ \\
\\
$\gamma_{RV}$ [m/s] & \multicolumn{3}{c }{$46575.7_{-13}^{+5.8}$} & $46922.6_{-6.6}^{+6.8}$ \\
\\
jitter HARPS-N [m/s] & \multicolumn{3}{c }{$2.08 \pm 0.29$} & $5.20_{-0.64}^{+0.72}$ \\
\\
jitter HIRES [m/s] & \multicolumn{3}{c }{-} & $8.5_{-2.0}^{+3.0}$ \\
\\
jitter HDS [m/s] & \multicolumn{3}{c }{-} & $6.6_{-2.1}^{+3.0}$ \\
\\
\hline
\end{tabular}\\
\vspace{0.2cm}
Physical and orbital parameters for the planets of both targets derived from RV fitting. The value of the mass of XO-2Nb is not the $m\sin i$ but rather the true mass. The reported values are the medians of the resulting distributions and the error bars are the 15th and 84th percentile, respectively. The parameter $T_0$ is the time of the periastron passage calculated with reference time BJD $= 2458842$. \textit{(1):} Values derived using priors obtained from the TESS light curves of XO-2N.
\end{table*}

\begin{table}
\centering
\caption{\label{tabgp} Parameters obtained for the multidimensional GP of XO-2N.}
\begin{tabular}{ c c }
\hline
\\
 Parameter & Value \\
\\
\hline
\\
P [d] & $3378_{-189}^{+284}$ \\
\\
$V_r$ [m/s] & 0 (fixed) \\
\\
$V_c$ [m/s] & $12.0_{-4.1}^{+4.6}$ \\
 \\
$L_c$ & $0.0270_{-0.0092}^{+0.012}$ \\
\\
$B_r$ & $1.37_{-0.76}^{+1.10}$ \\
\\
$B_c$ & $0.0070_{-0.0031}^{+0.0041}$ \\
\\
\hline
\end{tabular}\\
\vspace{0.2cm}
\end{table}

\subsection{Detection limits for additional planets}

\begin{figure}
    \centering
    \subfloat[][]
    {\includegraphics[width=0.45\textwidth]{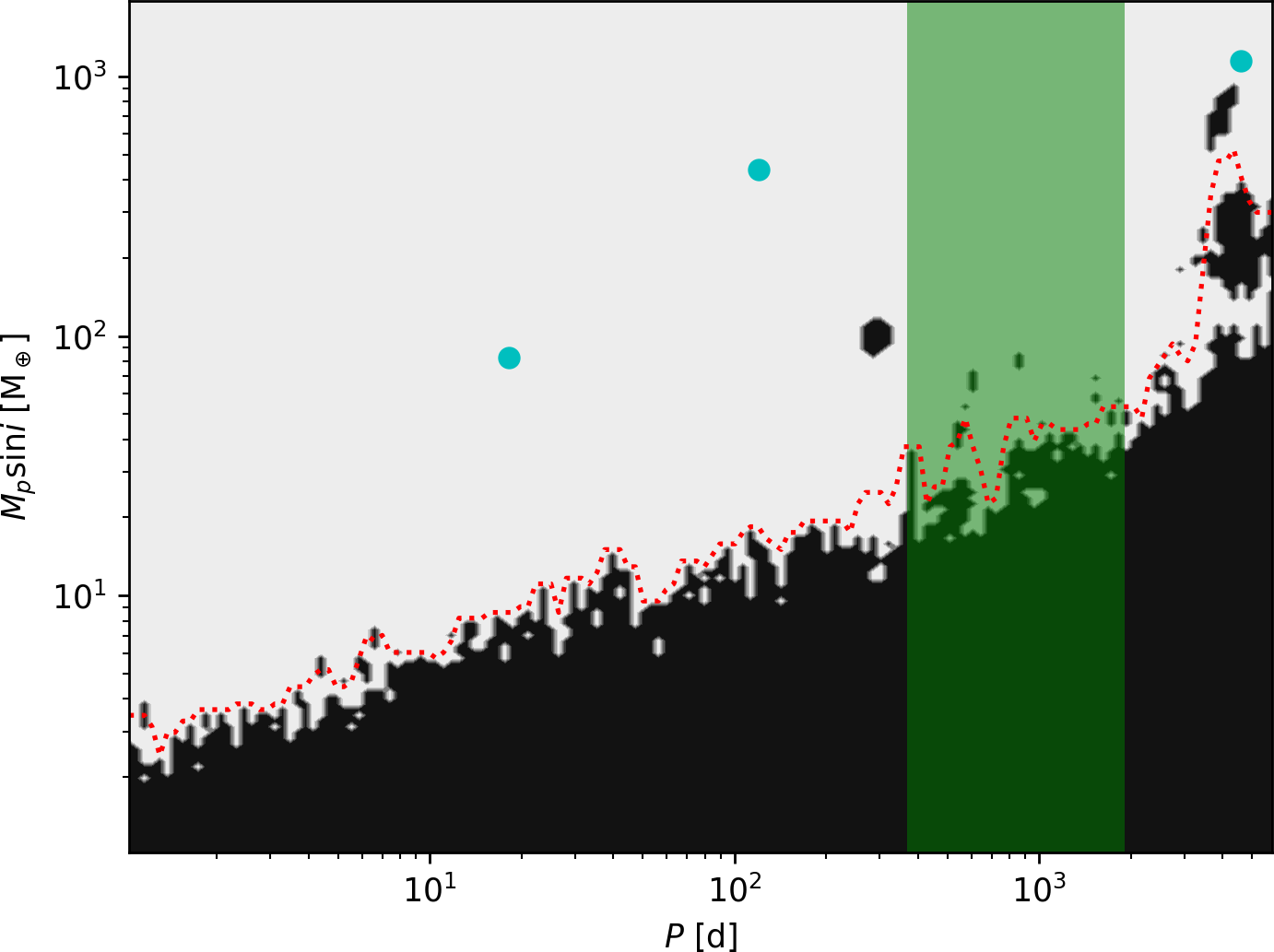}
    \label{fig:detection_maps:xo2s}}\\
    \subfloat[][]
    {\includegraphics[width=0.45\textwidth]{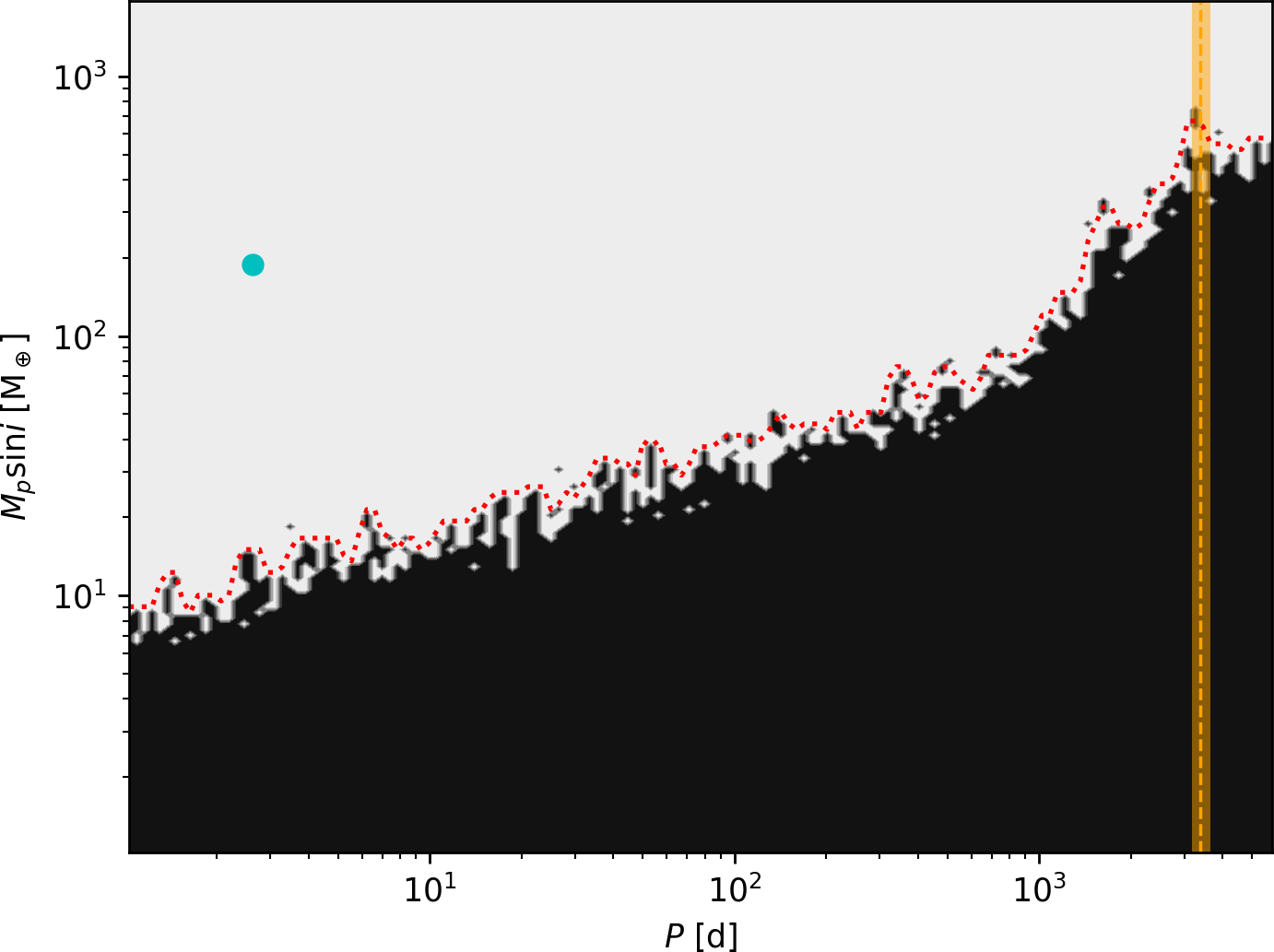}
    \label{fig:detection_maps:xo2n}}
    \caption{Detection function maps of the RV time series of XO-2S (upper panel) and XO-2N (lower panel). The white part corresponds to the area in the period-minimum mass space where additional signals could be detected if present in the data, while the black region corresponds to the area where the detection probability is negligible. The red dotted line marks the $95\%$ detection threshold. The cyan dots show the positions of the planets orbiting the two stars. The green shaded area in the upper panel denotes the stability region discussed in Sect. \ref{sec:stability}. The orange dashed line and shaded band in the lower panel mark the period and 1$\sigma$ uncertainty of the activity cycle as modeled in Sect. \ref{sec:cycle_GP}.}
    \label{fig:detection_maps}
\end{figure}

We investigate the presence of additional undetected companions, deriving the detection limits from the RV time series of the two stars. We adopted the Bayesian approach described in \citet{Pinamontietal2022} to obtain an unbiased estimate of the detectability function and detection limits of the collected RV data.
We applied this technique to the full RV time series, including all the periodic signals discussed in the previous sections, the Keplerian signals of the planetary companions, and the GP model of the XO-2N magnetic cycle. For the GP modeling of the activity cycle, we adopted the best-fit values and uncertainties for the multidimensional fit, listed in Table \ref{tabgp}, as priors for the GP hyper-parameters, as the activity indices were not used in the computation of the detection function.
Fig. \ref{fig:detection_maps} shows the resulting detection function maps.\\

In Fig. \ref{fig:detection_maps:xo2s} we can see the detection map of XO2-S: in the stability region discussed in Sect. \ref{sec:stability}, highlighted in green, the minimum mass detection threshold is $m \sin{i}_\text{thr} = 32^{+11}_{-12}$ M$_\oplus$. This shows that indeed a low-mass planet could exist in that region and remain undetected. Since the time span covered by our data is already longer than the period of a potential planet in that part of the parameter space, getting more data would not be very useful in this sense unless individual errors are reduced, the sampling significantly increased, or the number of additional data is large. It is also worth noticing that in the innermost stable regions mentioned before ($a < 0.1$ au), the collected RV time series exclude the presence of any additional companions with minimum mass $m \sin{i} \gtrsim 3$ M$_\oplus$. Thus, the only thing we cannot detect at such short distances is an Earth twin. 

From the detection map of XO-2N, shown in Fig. \ref{fig:detection_maps:xo2n}, we can see that the sensitivity of the RV time series is much lower than for the stellar companion, which is mainly due to the lower number of RV data and the sparser sampling of the observations of XO-2N with respect to XO-2S. It is also worth noticing that the sensitivity decreases significantly around the period of the activity cycle, highlighted in red, showing how the presence of a stellar magnetic cycle can obstruct the detection of planetary companions of similar periods (as would be the case of Jupiter and the Solar cycle), even when the effect of the cycle is modeled with advanced techniques such as GP regression. In any case, we conclude that, at present, XO-2N appears to be a single-planet system, except for potential low-mass objects that might escape our search. In particular, the minimum mass detection threshold is $m \sin{i}_\text{thr} = 11.1^{+4.7}_{-2.1}$ M$_\oplus$ in the 1-10 d interval, $m \sin{i}_\text{thr} = 23.7^{+8.4}_{-7.1}$ M$_\oplus$ in the 10-100 d range, and $m \sin{i}_\text{thr} = 51^{+18}_{-17}$ M$_\oplus$ in the 100-1000 d region.

\subsection{Limits on the masses of possible outer companions from Gaia and direct imaging}
\label{astrometry_imaging}

Further constraints on the mass of the detected planets and on the presence of additional companions
can in principle be derived from high-precision absolute astrometry. However, the XO-2 star system was not observed by Hipparcos, ruling out the possibility
of exploiting the sensitive Gaia-Hipparcos proper motion anomaly \citep{kervella2022}
and at $d\simeq150$ pc the constraints from Gaia alone are not very informative. The Gaia DR3 time span covers only $\sim20\%$ of the orbit of XO-2S\,d. The Renormalized Unit Weight Error (RUWE) statistics \citep{Lindegren2021} can be used to find indications of a statistically significant departure from a good single-star fit, typically when RUWE $\gtrsim 1.4$. For XO-2S, the Gaia archive reports RUWE $=0.88$, providing no indication of the presence of orbital motion. Following the approach described in \citet{Blanco-Pozo2023} and \citet{Kunimoto2023}, we estimated that at the separation of 5.4 au there is a probability $>90\%$ of obtaining RUWE $>0.88$ for true companion masses $\gtrsim60$ $M_\mathrm{Jup}$, ruling out only inclination angles $\lesssim 4^\circ$. 

Additional constraints at wider separation can be inferred by the proper motion differences.
Considering the individual components, there are marginally significant differences between proper motion in right ascension from Gaia DR3 and the one from Tycho2 and PPMXL catalogs, while UCAC5 agrees within errors. However, we noticed that such differences are similar in magnitude and sign for both components, suggesting that they are mostly due to systematic errors in the pre-Gaia catalogs, as found by, for example, \cite{lindegren2016} and \cite{shi2019}, rather than to the presence of additional bodies or to the orbital motion of the binary.

From the proper motion differences in Gaia DR3, we derive an instantaneous (at Gaia DR3 epoch) velocity
difference of 171$\pm$15 m/s and 3$\pm$15 m/s in right ascension and declination directions, respectively.
The velocity difference along the line of sight is instead derived from the difference of the center of mass velocities from the orbital fits in Sect. \ref{sec:fit_xo2s}   -\ref{sec:cycle_GP}. This amounts to 346$\pm$12 m/s \footnote{The uncertainty here includes the measurement uncertainty only. In general, when comparing the absolute RV of two stars, differences in convective blueshift and gravitational redshift are significant as well as color-dependent RV zero point errors, due to, for example, cross-correlation mask mismatch. However, these sources of errors are expected to be very small in the comparison of XO-2N and XO-2S, considering the very similar stellar properties and the common instrument setup. }.
Therefore, it results that the total velocity difference between the components is 386$\pm$24 m/s, for
a projected separation of 4685 au on the plane of the sky.
The physical separation along the line of sight remains unknown, as the individual parallaxes agree to better than one sigma, but a broad range of values is compatible with a bound orbit for XO-2N and XO-2S. We conclude that the observed differences between the components in RV and proper motion are fully consistent with those expected from orbital motion in a bound orbit, and the presence of additional massive companions is not required.

Finally, we note that both Gaia and dedicated imaging observations of XO-2N (\citealt{daemgen2009} and \citealt{ngo2015}) did not reveal additional
sources close to the star, ruling out any stellar companion at projected separations 
from 150 to 1500 au, and with detection limits of about 0.28 \msun\ at 30 au.
For XO-2S, there are no imaging observations on the target; relying on detection limits for close companions from Gaia \citep{brandeker2019}, companions more massive than 0.5, 0.25, 0.10 \msun\ are ruled out at projected separations of 180, 300, and 700 au, respectively.

\subsection{Revisiting the origin of the elemental abundances difference}



\cite{ramirez2015} and \cite{biazzo2015} discussed about the significant difference of the XO-2N abundance relative to XO-2S and the trends with the condensation temperature ($T_{\rm cond}$). They interpreted their results as the signature of the ingestion of material by XO-2N or depletion in XO-2S due to the locking of heavy elements by planetary companions. As described in the previous sections, we now have evidence for a new massive planet in the XO-2S system, while we ruled out a Keplerian origin for the long-term RV variations of XO-2N, excluding additional massive planets within several au. This leads to a large difference between the two stars in terms of total hosted planetary mass. In particular, we have a total minimum planetary mass of 5.35 \mjup for XO-2S and a true mass of 0.6 \mjup for XO-2N, so the ratio is approximately a factor of ten. In addition, since in the last years, some steps have been made in the field of abundance differences in binary components, here in the following we update possible implications related to the difference observed in the XO-2 system.

\begin{figure}[h]
   \centering
   \includegraphics[width = 90 mm]{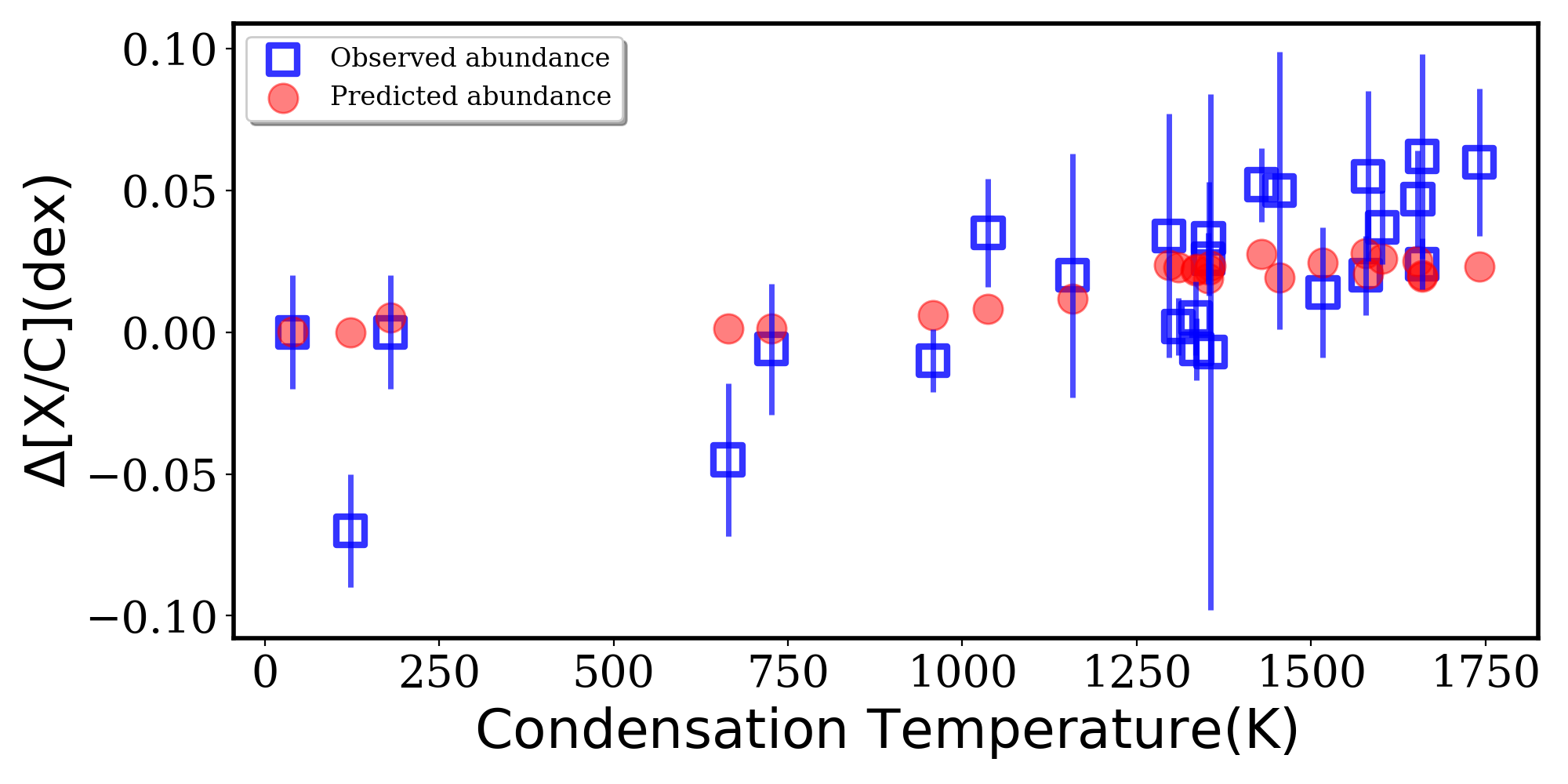}
   \includegraphics[width = 90 mm]{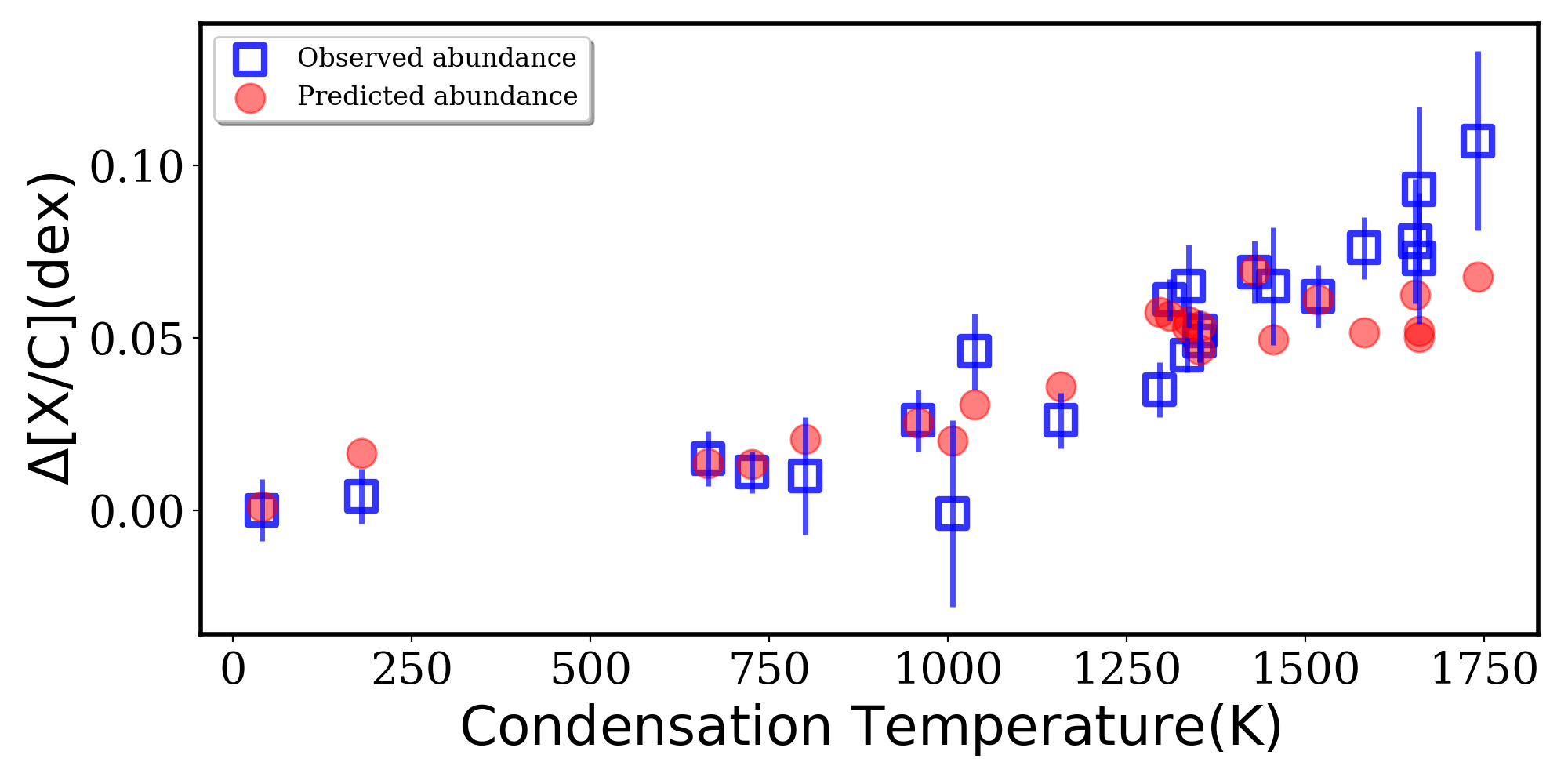}

      \caption{Comparison of the abundances of XO-2N relative to XO-2S (open blue squares) as a function of the dust condensation temperature, and the predicted abundances (red filled circles) estimated from a planetary engulfment scenario. Squares in the upper and lower panels represent the results by \cite{biazzo2015} and \cite{ramirez2015}, respectively. 
             }
         \label{fig:elemental_abundance_diff}
   \end{figure}

In order to better quantify and possibly validate the scenario of planetary material ingestion, we used the code \texttt{terra}\footnote{\url{https://github.com/ramstojh/terra}} (\citealt{yanagalarzaetal2016}), which first computes the convective mass of a solar-type star, and then estimates the mass of the rocky material missing in the convective zone. The convective mass is calculated considering the mass and the metallicity of the star and adopting the Yale isochrones of stellar evolution by \cite{Yietal2001}. Once the convective mass is obtained, the code computes the amount of rocky material necessary to reproduce the observed abundance pattern by adding meteoritic (\citealt{WassonKallemeyn1988}) and terrestrial (\citealt{Allegreetal2001}) abundance patterns into the convective zone of the star (see \citealt{yanagalarzaetal2016} and \citealt{yanagalarzaetal2021} for more details). Considering the stellar mass and iron abundance found by \cite{damasso2015}, the differential abundance pattern found by \cite{biazzo2015} can be explained by a rocky planet engulfment of $\sim$8.5\,$M_\oplus$, which is a mixture of $\sim$2\,$M_\oplus$ of chondrite-like composition and $\sim$6.5\,$M_\oplus$ of Earth-like composition. As can be seen in Fig.\,\ref{fig:elemental_abundance_diff}, the predicted (red filled circles) differential abundances seem to be smaller than those observed (open blue squares) for the elements with condensation temperature higher than $\sim$1500 K. Similar behavior happens also applying the same procedure to the results of \cite{ramirez2015} (see lower panel of Fig.\,\ref{fig:elemental_abundance_diff}), thus implying that possibly other scenarios, such as the locking of heavy elements around XO-2S, besides planet accretion onto XO-2N, could be the reason of the observed trend. However, to our knowledge, there are still no theoretical models predicting the amount of locking of heavy elements due to the presence of planetary companions. The scenario is also complicated by the presence of a multiple system around the S component with no transiting planets which prevents the determination of the planet density, whose value could be related to the content of heavy elements (see, e.g., \citealt{Biazzoetal2022}, and references therein). At the same time, if the planet engulfment scenario is correct, the rotation of XO-2N should be faster than XO-2S (see \citealt{Oetjensetal2020}). However, there is no significant difference in the homogeneously derived projected rotational velocity of the components, resulting in consistent $v \sin i$ within the errors (see \citealt{Biazzoetal2022}). This finding does not invalidate the hypothesis of planet ingestion because the planetary engulfment could have happened when the star was young, implying that the increase in angular momentum due to the planet has already been lost. Recent models indeed claim that engulfment events last no longer than $\sim 2$\, Gyr and that the detection of possible engulfment is rare to be detected in systems that are several Gyr old (see \citealt{behmardetal2023}). 

Interestingly, the absolute differences in [Fe/H] ($\sim 0.055$\,dex), the absolute value of the slope of the $T_{\rm cond}$ trend ($\sim 4.7\times 10^{-5}$\, dex K$^{-1}$), and the binary separation ($4600$\,au) seem to be in line with the recent trends proposed by \cite{liuetal2021} (see their Figs.\,15 and 16). The authors claim that the possible dependence of the [Fe/H] difference and the $T_{\rm cond}$ slope on the binary separation could indicate that the dynamical history of binaries has a potential impact on the process of planet formation. Although still speculative, it is possible that the architecture of planetary systems might be affected in binaries to some level, since the dynamical interaction and the binary separation are relevant. 

Another possible interpretation of the observed trends between the differential elemental abundances and the condensation temperatures could be offered by the framework of the gas-dust segregation scenario (\citealt{Gaidos2015}). Based on this idea, \cite{BoothOwen2020} developed an evolutionary model to test if, during the giant planet formation, the gas-dust segregation process operating across a protoplanetary disk could produce chemical imprints. The authors demonstrated that a distant forming giant planet can open gaps in the disk, creating a pressure trap and allowing the gas to be accreted onto the protostar. In contrast, if the planet forms early enough when the disk is still massive, the planet can trap a substantial mass of dust exterior to its orbit preventing the dust from accreting onto the star in contrast to the gas. This could result in refractory element deficiency of $5-15$\,\%, with the larger values occurring for conditions favoring giant planet formation around more massive and longer-lived disks. Taking into account the results by the authors and the mean elemental abundance difference between the N and the S components of $+0.067\pm0.032$\, dex (\citealt{biazzo2015}), the lack of refractory elements of the S component relative to the N companion maybe was produced through a gas-dust segregation mechanism at the early formation of its distant giant planet XO-2Sd ($a \sim 5.5$\,au, $m \sin i \sim 3.7\, M_{\rm J}$; see Table\,\ref{tabparam}). However, the statistically significant trend observed with the condensation temperature for the individual elements cannot still be explained with this model.

On the observational side, we mention that 
\cite{ramirez2015} and \cite{biazzo2015} also found hints of break-in $T_{\rm cond}$ between volatile and refractory elements at $\sim 800$\, K possibly related to the distance where the planets form along the protoplanetary disk. Theoretical simulations by \cite{bitschetal2018} showed that a planet formed inside the H$_2$O/CO ice line can produce a much larger difference in its host stars' surface abundance ratios ([Fe/O] and [Fe/C]) than for planets formed out of the H$_2$O/CO ice line. This is because at the disk location interior to the snow lines, the accreted core is volatile-depleted and the host stars would reveal a high abundance of volatile elements compared to the refractory ones. Following this hypothesis, the higher values of these elemental ratios found for XO-2N ([Fe/O]=0.050, [Fe/C]=-0.040) compared to XO-2S ([Fe/O]=-0.045, [Fe/C]=-0.045) could be indicative of the possible different location of the forming planets along the protoplanetary disks around the binary components.

The evaluation of these scenarios remains rather speculative at this stage. Further inputs might come on one side from additional modeling of the impact of planet formation processes on star elemental abundances, on the other side from a better understanding of the most probable formation sites and evolutionary histories of the planets in the system, through, for instance, population synthesis of the two systems \citep{mordasini2018} and more complete characterization of the transiting planet XO-2N b. Indeed, the only species significantly detected so far are potassium \citep{sing2011} and sodium \citep{sing2012}. \cite{pearson2019} spectrally resolved sodium and were able to set a slightly substellar lower limit to the abundance of this element by resolving the wings of the doublet, but unfortunately - because alkali elements are susceptible to ionization, and undetected clouds could bias the inferred abundances - it is not possible to draw firm conclusions on the overall atmospheric enrichment of XO-2N b based on this measurement.

\section{Conclusion}
\label{sec:conclusion}

In this paper, we analyzed the XO-2 system that had already been shown to be a unique benchmark for studying the stochasticity of planet formation. Indeed, it is a  binary system with nearly identical components, apart from a small difference in chemical abundances, both with a planetary system but with markedly different characteristics. Greatly extending the time coverage of our HARPS-N RV monitoring, we were able to further confirm the known planets and show evidence for an additional massive long-period giant planet around XO-2S in a Jupiter-like orbit. In addition, we found no evidence for the transits of its two inner planets, ruling out inclinations close to an edge-on configuration. Furthermore, we showed that, even for highly inclined orbits, the system remains stable and there are fairly broad stability regions where low-mass planets could be present but undetectable at present. 

On the other hand, for XO-2N we showed that the additional long-term RV signal is likely due to a Solar-like activity cycle rather than a second planet. Therefore, the known planet XO-2N b appears to be a single transiting hot Jupiter, although detection limits for additional companions
are poorer than those of XO-2S because of the lower amount of data and the higher effect of stellar activity.
Astrometry and direct imaging datasets have limited sensitivity but they do not support the presence of additional massive companions at large separations.

With these results, the known planetary system of XO-2N has a total mass of $0.59$ \mjup, while that of XO-2S has a minimum mass of $5.35$ \mjup. This means that, even though the two stars are almost physically identical, the more metallic one has one order of magnitude less planetary mass than the other. This might indicate that mechanisms related to the segregation of dust grains in planetary cores and atmospheres could be at work, although more detailed work, especially on the modeling side, is needed to reach firm conclusions.
In any case, the XO-2 system remains and becomes, even more after the new results of this paper, a unique laboratory to understand the diversity of the outcomes of the planet formation process.




\begin{acknowledgements}
We acknowledge the financial support from the agreement ASI-INAF n.2018-16-HH.0 \\
This work has made use of data from the European Space Agency (ESA) mission {\it Gaia} (\url{https://www.cosmos.esa.int/gaia}), processed by the {\it Gaia} Data Processing and Analysis Consortium (DPAC,
\url{https://www.cosmos.esa.int/web/gaia/dpac/consortium}). Funding for the DPAC has been provided by national institutions, in particular, the institutions participating in the {\it Gaia} Multilateral Agreement.
We acknowledge the use of public TESS data from pipelines at the TESS Science Office and at the TESS Science Processing Operations Center. Resources supporting this work were provided by the NASA High-End Computing (HEC) Program through the NASA Advanced Supercomputing (NAS) Division at Ames Research Center for the production of the SPOC data products.
\end{acknowledgements}

\bibliography{biblio.bib} 

\begin{thebibliography}{75}
\expandafter\ifx\csname natexlab\endcsname\relax\def\natexlab#1{#1}\fi

\bibitem[{{All{\`e}gre} {et~al.}(2001){All{\`e}gre}, {Manh{\`e}s}, \&
  {Lewin}}]{Allegreetal2001}
{All{\`e}gre}, C., {Manh{\`e}s}, G., \& {Lewin}, {\'E}. 2001, Earth and
  Planetary Science Letters, 185, 49

\bibitem[{{Behmard} {et~al.}(2023){Behmard}, {Dai}, {Brewer}, {Berger}, \&
  {Howard}}]{behmardetal2023}
{Behmard}, A., {Dai}, F., {Brewer}, J.~M., {Berger}, T.~A., \& {Howard}, A.~W.
  2023, \mnras, 521, 2969

\bibitem[{{Benatti} {et~al.}(2017){Benatti}, {Desidera}, {Damasso},
  {Malavolta}, {Lanza}, {Biazzo}, {Bonomo}, {Claudi}, {Marzari}, {Poretti},
  {Gratton}, {Micela}, {Pagano}, {Piotto}, {Sozzetti}, {Boccato}, {Cosentino},
  {Covino}, {Maggio}, {Molinari}, {Smareglia}, {Affer}, {Andreuzzi},
  {Bignamini}, {Borsa}, {di Fabrizio}, {Esposito}, {Martinez Fiorenzano},
  {Messina}, {Giacobbe}, {Harutyunyan}, {Knapic}, {Maldonado}, {Masiero},
  {Nascimbeni}, {Pedani}, {Rainer}, {Scandariato}, \& {Silvotti}}]{benatti2017}
{Benatti}, S., {Desidera}, S., {Damasso}, M., {et~al.} 2017, \aap, 599, A90

\bibitem[{{Biazzo} {et~al.}(2022){Biazzo}, {D'Orazi}, {Desidera}, {Turrini},
  {Benatti}, {Gratton}, {Magrini}, {Sozzetti}, {Baratella}, {Bonomo}, {Borsa},
  {Claudi}, {Covino}, {Damasso}, {Di Mauro}, {Lanza}, {Maggio}, {Malavolta},
  {Maldonado}, {Marzari}, {Micela}, {Poretti}, {Vitello}, {Affer}, {Bignamini},
  {Carleo}, {Cosentino}, {Fiorenzano}, {Giacobbe}, {Harutyunyan}, {Leto},
  {Mancini}, {Molinari}, {Molinaro}, {Nardiello}, {Nascimbeni}, {Pagano},
  {Pedani}, {Piotto}, {Rainer}, \& {Scandariato}}]{Biazzoetal2022}
{Biazzo}, K., {D'Orazi}, V., {Desidera}, S., {et~al.} 2022, \aap, 664, A161

\bibitem[{{Biazzo} {et~al.}(2015){Biazzo}, {Gratton}, {Desidera}, {Lucatello},
  {Sozzetti}, {Bonomo}, {Damasso}, {Gandolfi}, {Affer}, {Boccato}, {Borsa},
  {Claudi}, {Cosentino}, {Covino}, {Knapic}, {Lanza}, {Maldonado}, {Marzari},
  {Micela}, {Molaro}, {Pagano}, {Pedani}, {Pillitteri}, {Piotto}, {Poretti},
  {Rainer}, {Santos}, {Scandariato}, \& {Zanmar Sanchez}}]{biazzo2015}
{Biazzo}, K., {Gratton}, R., {Desidera}, S., {et~al.} 2015, \aap, 583, A135

\bibitem[{{Bitsch} {et~al.}(2018){Bitsch}, {Forsberg}, {Liu}, \&
  {Johansen}}]{bitschetal2018}
{Bitsch}, B., {Forsberg}, R., {Liu}, F., \& {Johansen}, A. 2018, \mnras, 479,
  3690

\bibitem[{{Blanco-Pozo} {et~al.}(2023){Blanco-Pozo}, {Perger}, {Damasso},
  {Anglada Escud{\'e}}, {Ribas}, {Baroch}, {Caballero}, {Cifuentes}, {Jeffers},
  {Lafarga}, {Kaminski}, {Kaur}, {Nagel}, {Perdelwitz}, {P{\'e}rez-Torres},
  {Sozzetti}, {Vigan{\`o}}, {Amado}, {Andreuzzi}, {B{\'e}jar}, {Brown}, {Del
  Sordo}, {Dreizler}, {Galad{\'\i}-Enr{\'\i}quez}, {Hatzes}, {K{\"u}rster},
  {Lanza}, {Melis}, {Molinari}, {Montes}, {Murgia}, {Pall{\'e}},
  {Pe{\~n}a-Mo{\~n}ino}, {Perrodin}, {Pilia}, {Poretti}, {Quirrenbach},
  {Reiners}, {Schweitzer}, {Zapatero Osorio}, \&
  {Zechmeister}}]{Blanco-Pozo2023}
{Blanco-Pozo}, J., {Perger}, M., {Damasso}, M., {et~al.} 2023, \aap, 671, A50

\bibitem[{{Booth} \& {Owen}(2020)}]{BoothOwen2020}
{Booth}, R.~A. \& {Owen}, J.~E. 2020, \mnras, 493, 5079

\bibitem[{{Brandeker} \& {Cataldi}(2019)}]{brandeker2019}
{Brandeker}, A. \& {Cataldi}, G. 2019, \aap, 621, A86

\bibitem[{{Burke} {et~al.}(2007){Burke}, {McCullough}, {Valenti},
  {Johns-Krull}, {Janes}, {Heasley}, {Summers}, {Stys}, {Bissinger}, {Fleenor},
  {Foote}, {Garc{\'\i}a-Melendo}, {Gary}, {Howell}, {Mallia}, {Masi}, {Taylor},
  \& {Vanmunster}}]{burke2007}
{Burke}, C.~J., {McCullough}, P.~R., {Valenti}, J.~A., {et~al.} 2007, \apj,
  671, 2115

\bibitem[{{Butler} {et~al.}(2017){Butler}, {Vogt}, {Laughlin}, {Burt},
  {Rivera}, {Tuomi}, {Teske}, {Arriagada}, {Diaz}, {Holden}, \&
  {Keiser}}]{Butler2017}
{Butler}, R.~P., {Vogt}, S.~S., {Laughlin}, G., {et~al.} 2017, \aj, 153, 208

\bibitem[{{Cincotta} \& {Sim{\'o}}(2000)}]{MEGNO1}
{Cincotta}, P.~M. \& {Sim{\'o}}, C. 2000, \aaps, 147, 205

\bibitem[{{Cosentino} {et~al.}(2012){Cosentino}, {Lovis}, {Pepe}, {Collier
  Cameron}, {Latham}, {Molinari}, {Udry}, {Bezawada}, {Black}, {Born},
  {Buchschacher}, {Charbonneau}, {Figueira}, {Fleury}, {Galli}, {Gallie},
  {Gao}, {Ghedina}, {Gonzalez}, {Gonzalez}, {Guerra}, {Henry}, {Horne},
  {Hughes}, {Kelly}, {Lodi}, {Lunney}, {Maire}, {Mayor}, {Micela}, {Ordway},
  {Peacock}, {Phillips}, {Piotto}, {Pollacco}, {Queloz}, {Rice}, {Riverol},
  {Riverol}, {San Juan}, {Sasselov}, {Segransan}, {Sozzetti}, {Sosnowska},
  {Stobie}, {Szentgyorgyi}, {Vick}, \& {Weber}}]{Cosentino2012}
{Cosentino}, R., {Lovis}, C., {Pepe}, F., {et~al.} 2012, in Society of
  Photo-Optical Instrumentation Engineers (SPIE) Conference Series, Vol. 8446,
  Ground-based and Airborne Instrumentation for Astronomy IV, ed. I.~S.
  {McLean}, S.~K. {Ramsay}, \& H.~{Takami}, 84461V

\bibitem[{{Covino} {et~al.}(2013){Covino}, {Esposito}, {Barbieri}, {Mancini},
  {Nascimbeni}, {Claudi}, {Desidera}, {Gratton}, {Lanza}, {Sozzetti}, {Biazzo},
  {Affer}, {Gandolfi}, {Munari}, {Pagano}, {Bonomo}, {Collier Cameron},
  {H{\'e}brard}, {Maggio}, {Messina}, {Micela}, {Molinari}, {Pepe}, {Piotto},
  {Ribas}, {Santos}, {Southworth}, {Shkolnik}, {Triaud}, {Bedin}, {Benatti},
  {Boccato}, {Bonavita}, {Borsa}, {Borsato}, {Brown}, {Carolo}, {Ciceri},
  {Cosentino}, {Damasso}, {Faedi}, {Mart{\'\i}nez Fiorenzano}, {Latham},
  {Lovis}, {Mordasini}, {Nikolov}, {Poretti}, {Rainer}, {Rebolo L{\'o}pez},
  {Scandariato}, {Silvotti}, {Smareglia}, {Alcal{\'a}}, {Cunial}, {Di
  Fabrizio}, {Di Mauro}, {Giacobbe}, {Granata}, {Harutyunyan}, {Knapic},
  {Lattanzi}, {Leto}, {Lodato}, {Malavolta}, {Marzari}, {Molinaro},
  {Nardiello}, {Pedani}, {Prisinzano}, \& {Turrini}}]{covino2013}
{Covino}, E., {Esposito}, M., {Barbieri}, M., {et~al.} 2013, \aap, 554, A28

\bibitem[{{Daemgen} {et~al.}(2009){Daemgen}, {Hormuth}, {Brandner}, {Bergfors},
  {Janson}, {Hippler}, \& {Henning}}]{daemgen2009}
{Daemgen}, S., {Hormuth}, F., {Brandner}, W., {et~al.} 2009, \aap, 498, 567

\bibitem[{{Damasso} {et~al.}(2015){Damasso}, {Biazzo}, {Bonomo}, {Desidera},
  {Lanza}, {Nascimbeni}, {Esposito}, {Scandariato}, {Sozzetti}, {Cosentino},
  {Gratton}, {Malavolta}, {Rainer}, {Gandolfi}, {Poretti}, {Zanmar Sanchez},
  {Ribas}, {Santos}, {Affer}, {Andreuzzi}, {Barbieri}, {Bedin}, {Benatti},
  {Bernagozzi}, {Bertolini}, {Bonavita}, {Borsa}, {Borsato}, {Boschin},
  {Calcidese}, {Carbognani}, {Cenadelli}, {Christille}, {Claudi}, {Covino},
  {Cunial}, {Giacobbe}, {Granata}, {Harutyunyan}, {Lattanzi}, {Leto},
  {Libralato}, {Lodato}, {Lorenzi}, {Mancini}, {Martinez Fiorenzano},
  {Marzari}, {Masiero}, {Micela}, {Molinari}, {Molinaro}, {Munari}, {Murabito},
  {Pagano}, {Pedani}, {Piotto}, {Rosenberg}, {Silvotti}, \&
  {Southworth}}]{damasso2015}
{Damasso}, M., {Biazzo}, K., {Bonomo}, A.~S., {et~al.} 2015, \aap, 575, A111

\bibitem[{{Desidera} {et~al.}(2014){Desidera}, {Bonomo}, {Claudi}, {Damasso},
  {Biazzo}, {Sozzetti}, {Marzari}, {Benatti}, {Gandolfi}, {Gratton}, {Lanza},
  {Nascimbeni}, {Andreuzzi}, {Affer}, {Barbieri}, {Bedin}, {Bignamini},
  {Bonavita}, {Borsa}, {Calcidese}, {Christille}, {Cosentino}, {Covino},
  {Esposito}, {Giacobbe}, {Harutyunyan}, {Latham}, {Lattanzi}, {Leto},
  {Lodato}, {Lovis}, {Maggio}, {Malavolta}, {Mancini}, {Martinez Fiorenzano},
  {Micela}, {Molinari}, {Mordasini}, {Munari}, {Pagano}, {Pedani}, {Pepe},
  {Piotto}, {Poretti}, {Rainer}, {Ribas}, {Santos}, {Scandariato}, {Silvotti},
  {Southworth}, \& {Zanmar Sanchez}}]{desidera2014}
{Desidera}, S., {Bonomo}, A.~S., {Claudi}, R.~U., {et~al.} 2014, \aap, 567, L6

\bibitem[{{Eggenberger} \& {Udry}(2010)}]{eggenberger2010}
{Eggenberger}, A. \& {Udry}, S. 2010, in EAS Publications Series, Vol.~41, EAS
  Publications Series, ed. T.~{Montmerle}, D.~{Ehrenreich}, \& A.~M.
  {Lagrange}, 27--75

\bibitem[{{Foreman-Mackey} {et~al.}(2013){Foreman-Mackey}, {Hogg}, {Lang}, \&
  {Goodman}}]{foreman2013}
{Foreman-Mackey}, D., {Hogg}, D.~W., {Lang}, D., \& {Goodman}, J. 2013, \pasp,
  125, 306

\bibitem[{{Gaidos}(2015)}]{Gaidos2015}
{Gaidos}, E. 2015, \apj, 804, 40

\bibitem[{{Galarza} {et~al.}(2021){Galarza}, {L{\'o}pez-Valdivia},
  {Mel{\'e}ndez}, \& {Lorenzo-Oliveira}}]{yanagalarzaetal2021}
{Galarza}, J.~Y., {L{\'o}pez-Valdivia}, R., {Mel{\'e}ndez}, J., \&
  {Lorenzo-Oliveira}, D. 2021, \apj, 922, 129

\bibitem[{{Gelman} \& {Rubin}(1992)}]{gelman1992}
{Gelman}, A. \& {Rubin}, D.~B. 1992, Statistical Science, 7, 457

\bibitem[{{Gomes da Silva} {et~al.}(2018){Gomes da Silva}, {Figueira},
  {Santos}, \& {Faria}}]{daSilva2018}
{Gomes da Silva}, J., {Figueira}, P., {Santos}, N., \& {Faria}, J. 2018, The
  Journal of Open Source Software, 3, 667

\bibitem[{{Gomes da Silva} {et~al.}(2021){Gomes da Silva}, {Santos},
  {Adibekyan}, {Sousa}, {Campante}, {Figueira}, {Bossini}, {Delgado-Mena},
  {Monteiro}, {de Laverny}, {Recio-Blanco}, \& {Lovis}}]{dasilva2021}
{Gomes da Silva}, J., {Santos}, N.~C., {Adibekyan}, V., {et~al.} 2021, \aap,
  646, A77

\bibitem[{{Go{\'z}dziewski} {et~al.}(2001){Go{\'z}dziewski}, {Bois},
  {Maciejewski}, \& {Kiseleva-Eggleton}}]{MEGNO2}
{Go{\'z}dziewski}, K., {Bois}, E., {Maciejewski}, A.~J., \&
  {Kiseleva-Eggleton}, L. 2001, \aap, 378, 569

\bibitem[{{Grunblatt} {et~al.}(2015){Grunblatt}, {Howard}, \&
  {Haywood}}]{Grunblatt2015}
{Grunblatt}, S.~K., {Howard}, A.~W., \& {Haywood}, R.~D. 2015, \apj, 808, 127

\bibitem[{{Hippke} \& {Heller}(2019)}]{hippke2019}
{Hippke}, M. \& {Heller}, R. 2019, \aap, 623, A39

\bibitem[{Husser {et~al.}(2013)Husser, {Wende-von Berg}, Dreizler, Homeier,
  Reiners, Barman, \& Hauschildt}]{Husser2013}
Husser, T.-O., {Wende-von Berg}, S., Dreizler, S., {et~al.} 2013, A{\&}A, 553,
  A6

\bibitem[{{Ivshina} \& {Winn}(2022)}]{ivshina2022}
{Ivshina}, E.~S. \& {Winn}, J.~N. 2022, \apjs, 259, 62

\bibitem[{Jenkins {et~al.}(2016)Jenkins, Twicken, McCauliff, Campbell,
  Sanderfer, Lung, Mansouri-Samani, Girouard, Tenenbaum, Klaus, Smith,
  Caldwell, Chacon, Henze, Heiges, Latham, Morgan, Swade, Rinehart, \&
  Vanderspek}]{jenkins2016}
Jenkins, J.~M., Twicken, J.~D., McCauliff, S., {et~al.} 2016, in Software and
  Cyberinfrastructure for Astronomy IV, ed. G.~Chiozzi \& J.~C. Guzman, Vol.
  9913, International Society for Optics and Photonics (SPIE), 99133E

\bibitem[{{Jones} {et~al.}(2006){Jones}, {Butler}, {Tinney}, {Marcy}, {Carter},
  {Penny}, {McCarthy}, \& {Bailey}}]{jones2006}
{Jones}, H. R.~A., {Butler}, R.~P., {Tinney}, C.~G., {et~al.} 2006, \mnras,
  369, 249

\bibitem[{{Kane}(2007)}]{kane2007}
{Kane}, S.~R. 2007, \mnras, 380, 1488

\bibitem[{Kass \& Raftery(1995)}]{kass1995}
Kass, R.~E. \& Raftery, A.~E. 1995, Journal of the American Statistical
  Association, 90, 773

\bibitem[{{Kervella} {et~al.}(2022){Kervella}, {Arenou}, \&
  {Th{\'e}venin}}]{kervella2022}
{Kervella}, P., {Arenou}, F., \& {Th{\'e}venin}, F. 2022, \aap, 657, A7

\bibitem[{{Kreidberg}(2015)}]{kreidberg2015}
{Kreidberg}, L. 2015, \pasp, 127, 1161

\bibitem[{{Kunimoto} {et~al.}(2023){Kunimoto}, {Vanderburg}, {Huang}, {Davis},
  {Affer}, {Cameron}, {Charbonneau}, {Cosentino}, {Damasso}, {Dumusque},
  {Fiorenzano}, {Ghedina}, {Haywood}, {Lienhard}, {L{\'o}pez-Morales}, {Mayor},
  {Pepe}, {Pinamonti}, {Poretti}, {Maldonado}, {Rice}, {Sozzetti}, {Wilson},
  {Udry}, {Baptista}, {Barkaoui}, {Becker}, {Benni}, {Bieryla}, {Bosch-Cabot},
  {Ciardi}, {Collins}, {Collins}, {Evans}, {Dupuy}, {Goliguzova}, {Guerra},
  {Kraus}, {Lissauer}, {Huber}, {Murgas}, {Palle}, {Quinn}, {Safonov},
  {Schwarz}, {Shporer}, {Stassun}, {Jenkins}, {Latham}, {Ricker}, {Seager},
  {Vanderspek}, {Winn}, {Essack}, {Lewis}, \& {Rose}}]{Kunimoto2023}
{Kunimoto}, M., {Vanderburg}, A., {Huang}, C.~X., {et~al.} 2023, \aj, 166, 7

\bibitem[{{Lanza} {et~al.}(2016){Lanza}, {Molaro}, {Monaco}, \&
  {Haywood}}]{lanza2016}
{Lanza}, A.~F., {Molaro}, P., {Monaco}, L., \& {Haywood}, R.~D. 2016, \aap,
  587, A103

\bibitem[{{Lanza} {et~al.}(2004){Lanza}, {Rodon{\`o}}, \& {Pagano}}]{lanza2004}
{Lanza}, A.~F., {Rodon{\`o}}, M., \& {Pagano}, I. 2004, \aap, 425, 707

\bibitem[{{Lightkurve Collaboration} {et~al.}(2018){Lightkurve Collaboration},
  {Cardoso}, {Hedges}, {Gully-Santiago}, {Saunders}, {Cody}, {Barclay}, {Hall},
  {Sagear}, {Turtelboom}, {Zhang}, {Tzanidakis}, {Mighell}, {Coughlin}, {Bell},
  {Berta-Thompson}, {Williams}, {Dotson}, \& {Barentsen}}]{lightkurve2018}
{Lightkurve Collaboration}, {Cardoso}, J.~V.~d.~M., {Hedges}, C., {et~al.}
  2018, {Lightkurve: Kepler and TESS time series analysis in Python},
  Astrophysics Source Code Library

\bibitem[{{Lindegren} {et~al.}(2021){Lindegren}, {Klioner}, {Hern{\'a}ndez},
  {Bombrun}, {Ramos-Lerate}, {Steidelm{\"u}ller}, {Bastian}, {Biermann}, {de
  Torres}, {Gerlach}, {Geyer}, {Hilger}, {Hobbs}, {Lammers}, {McMillan},
  {Stephenson}, {Casta{\~n}eda}, {Davidson}, {Fabricius}, {Gracia-Abril},
  {Portell}, {Rowell}, {Teyssier}, {Torra}, {Bartolom{\'e}}, {Clotet},
  {Garralda}, {Gonz{\'a}lez-Vidal}, {Torra}, {Abbas}, {Altmann}, {Anglada
  Varela}, {Balaguer-N{\'u}{\~n}ez}, {Balog}, {Barache}, {Becciani}, {Bernet},
  {Bertone}, {Bianchi}, {Bouquillon}, {Brown}, {Bucciarelli}, {Busonero},
  {Butkevich}, {Buzzi}, {Cancelliere}, {Carlucci}, {Charlot}, {Cioni},
  {Crosta}, {Crowley}, {del Peloso}, {del Pozo}, {Drimmel}, {Esquej}, {Fienga},
  {Fraile}, {Gai}, {Garcia-Reinaldos}, {Guerra}, {Hambly}, {Hauser},
  {Jan{\ss}en}, {Jordan}, {Kostrzewa-Rutkowska}, {Lattanzi}, {Liao}, {Licata},
  {Lister}, {L{\"o}ffler}, {Marchant}, {Masip}, {Mignard}, {Mints}, {Molina},
  {Mora}, {Morbidelli}, {Murphy}, {Pagani}, {Panuzzo}, {Pe{\~n}alosa Esteller},
  {Poggio}, {Re Fiorentin}, {Riva}, {Sagrist{\`a} Sell{\'e}s}, {Sanchez
  Gimenez}, {Sarasso}, {Sciacca}, {Siddiqui}, {Smart}, {Souami}, {Spagna},
  {Steele}, {Taris}, {Utrilla}, {van Reeven}, \& {Vecchiato}}]{Lindegren2021}
{Lindegren}, L., {Klioner}, S.~A., {Hern{\'a}ndez}, J., {et~al.} 2021, \aap,
  649, A2

\bibitem[{{Lindegren} {et~al.}(2016){Lindegren}, {Lammers}, {Bastian},
  {Hern{\'a}ndez}, {Klioner}, {Hobbs}, {Bombrun}, {Michalik}, {Ramos-Lerate},
  {Butkevich}, {Comoretto}, {Joliet}, {Holl}, {Hutton}, {Parsons},
  {Steidelm{\"u}ller}, {Abbas}, {Altmann}, {Andrei}, {Anton}, {Bach},
  {Barache}, {Becciani}, {Berthier}, {Bianchi}, {Biermann}, {Bouquillon},
  {Bourda}, {Br{\"u}semeister}, {Bucciarelli}, {Busonero}, {Carlucci},
  {Casta{\~n}eda}, {Charlot}, {Clotet}, {Crosta}, {Davidson}, {de Felice},
  {Drimmel}, {Fabricius}, {Fienga}, {Figueras}, {Fraile}, {Gai}, {Garralda},
  {Geyer}, {Gonz{\'a}lez-Vidal}, {Guerra}, {Hambly}, {Hauser}, {Jordan},
  {Lattanzi}, {Lenhardt}, {Liao}, {L{\"o}ffler}, {McMillan}, {Mignard}, {Mora},
  {Morbidelli}, {Portell}, {Riva}, {Sarasso}, {Serraller}, {Siddiqui}, {Smart},
  {Spagna}, {Stampa}, {Steele}, {Taris}, {Torra}, {van Reeven}, {Vecchiato},
  {Zschocke}, {de Bruijne}, {Gracia}, {Raison}, {Lister}, {Marchant},
  {Messineo}, {Soffel}, {Osorio}, {de Torres}, \& {O'Mullane}}]{lindegren2016}
{Lindegren}, L., {Lammers}, U., {Bastian}, U., {et~al.} 2016, \aap, 595, A4

\bibitem[{{Lissauer} {et~al.}(2014){Lissauer}, {Marcy}, {Bryson}, {Rowe},
  {Jontof-Hutter}, {Agol}, {Borucki}, {Carter}, {Ford}, {Gilliland}, {Kolbl},
  {Star}, {Steffen}, \& {Torres}}]{lissauer2014}
{Lissauer}, J.~J., {Marcy}, G.~W., {Bryson}, S.~T., {et~al.} 2014, \apj, 784,
  44

\bibitem[{{Liu} {et~al.}(2021){Liu}, {Bitsch}, {Asplund}, {Liu}, {Murphy},
  {Yong}, {Ting}, \& {Feltzing}}]{liuetal2021}
{Liu}, F., {Bitsch}, B., {Asplund}, M., {et~al.} 2021, \mnras, 508, 1227

\bibitem[{{Lovis} {et~al.}(2011){Lovis}, {Dumusque}, {Santos}, {Bouchy},
  {Mayor}, {Pepe}, {Queloz}, {S{\'e}gransan}, \& {Udry}}]{Lovis2011}
{Lovis}, C., {Dumusque}, X., {Santos}, N.~C., {et~al.} 2011, arXiv e-prints,
  arXiv:1107.5325

\bibitem[{{Malavolta} {et~al.}(2018){Malavolta}, {Mayo}, {Louden}, {Rajpaul},
  {Bonomo}, {Buchhave}, {Kreidberg}, {Kristiansen}, {Lopez-Morales}, {Mortier},
  {Vanderburg}, {Coffinet}, {Ehrenreich}, {Lovis}, {Bouchy}, {Charbonneau},
  {Ciardi}, {Collier Cameron}, {Cosentino}, {Crossfield}, {Damasso},
  {Dressing}, {Dumusque}, {Everett}, {Figueira}, {Fiorenzano}, {Gonzales},
  {Haywood}, {Harutyunyan}, {Hirsch}, {Howell}, {Johnson}, {Latham}, {Lopez},
  {Mayor}, {Micela}, {Molinari}, {Nascimbeni}, {Pepe}, {Phillips}, {Piotto},
  {Rice}, {Sasselov}, {S{\'e}gransan}, {Sozzetti}, {Udry}, \&
  {Watson}}]{Malavolta2018}
{Malavolta}, L., {Mayo}, A.~W., {Louden}, T., {et~al.} 2018, \aj, 155, 107

\bibitem[{{Malavolta} {et~al.}(2016){Malavolta}, {Nascimbeni}, {Piotto},
  {Quinn}, {Borsato}, {Granata}, {Bonomo}, {Marzari}, {Bedin}, {Rainer},
  {Desidera}, {Lanza}, {Poretti}, {Sozzetti}, {White}, {Latham}, {Cunial},
  {Libralato}, {Nardiello}, {Boccato}, {Claudi}, {Cosentino}, {Covino},
  {Gratton}, {Maggio}, {Micela}, {Molinari}, {Pagano}, {Smareglia}, {Affer},
  {Andreuzzi}, {Aparicio}, {Benatti}, {Bignamini}, {Borsa}, {Damasso}, {Di
  Fabrizio}, {Harutyunyan}, {Esposito}, {Fiorenzano}, {Gandolfi}, {Giacobbe},
  {Gonz{\'a}lez Hern{\'a}ndez}, {Maldonado}, {Masiero}, {Molinaro}, {Pedani},
  \& {Scandariato}}]{Malavolta2016}
{Malavolta}, L., {Nascimbeni}, V., {Piotto}, G., {et~al.} 2016, \aap, 588, A118

\bibitem[{{Mamajek} \& {Hillenbrand}(2008)}]{mamajek2008}
{Mamajek}, E.~E. \& {Hillenbrand}, L.~A. 2008, \apj, 687, 1264

\bibitem[{{Mantovan} {et~al.}(2022){Mantovan}, {Montalto}, {Piotto}, {Wilson},
  {Collier Cameron}, {Majidi}, {Borsato}, {Granata}, \&
  {Nascimbeni}}]{mantovan2022}
{Mantovan}, G., {Montalto}, M., {Piotto}, G., {et~al.} 2022, \mnras, 516, 4432

\bibitem[{{McLaughlin}(1924)}]{mclaughlin1924}
{McLaughlin}, D.~B. 1924, \apj, 60, 22

\bibitem[{{Mordasini}(2018)}]{mordasini2018}
{Mordasini}, C. 2018, in Handbook of Exoplanets, ed. H.~J. {Deeg} \& J.~A.
  {Belmonte}, 143

\bibitem[{{Narita} {et~al.}(2011){Narita}, {Hirano}, {Sato}, {Harakawa},
  {Fukui}, {Aoki}, \& {Tamura}}]{narita2011}
{Narita}, N., {Hirano}, T., {Sato}, B., {et~al.} 2011, \pasj, 63, L67

\bibitem[{{Neveu-VanMalle} {et~al.}(2014){Neveu-VanMalle}, {Queloz},
  {Anderson}, {Charbonnel}, {Collier Cameron}, {Delrez}, {Gillon}, {Hellier},
  {Jehin}, {Lendl}, {Maxted}, {Pepe}, {Pollacco}, {S{\'e}gransan}, {Smalley},
  {Smith}, {Southworth}, {Triaud}, {Udry}, \& {West}}]{neveu2014}
{Neveu-VanMalle}, M., {Queloz}, D., {Anderson}, D.~R., {et~al.} 2014, \aap,
  572, A49

\bibitem[{{Ngo} {et~al.}(2015){Ngo}, {Knutson}, {Hinkley}, {Crepp}, {Bechter},
  {Batygin}, {Howard}, {Johnson}, {Morton}, \& {Muirhead}}]{ngo2015}
{Ngo}, H., {Knutson}, H.~A., {Hinkley}, S., {et~al.} 2015, \apj, 800, 138

\bibitem[{{Noyes} {et~al.}(1984){Noyes}, {Hartmann}, {Baliunas}, {Duncan}, \&
  {Vaughan}}]{noyes1984}
{Noyes}, R.~W., {Hartmann}, L.~W., {Baliunas}, S.~L., {Duncan}, D.~K., \&
  {Vaughan}, A.~H. 1984, \apj, 279, 763

\bibitem[{{Oetjens} {et~al.}(2020){Oetjens}, {Carone}, {Bergemann}, \&
  {Serenelli}}]{Oetjensetal2020}
{Oetjens}, A., {Carone}, L., {Bergemann}, M., \& {Serenelli}, A. 2020, \aap,
  643, A34

\bibitem[{Parviainen \& Aigrain(2015)}]{Parviainen2015}
Parviainen, H. \& Aigrain, S. 2015, MNRAS, 453, 3821

\bibitem[{{Pearson} {et~al.}(2019){Pearson}, {Griffith}, {Zellem}, {Koskinen},
  \& {Roudier}}]{pearson2019}
{Pearson}, K.~A., {Griffith}, C.~A., {Zellem}, R.~T., {Koskinen}, T.~T., \&
  {Roudier}, G.~M. 2019, \aj, 157, 21

\bibitem[{{Pinamonti} {et~al.}(2022){Pinamonti}, {Sozzetti}, {Maldonado},
  {Affer}, {Micela}, {Bonomo}, {Lanza}, {Perger}, {Ribas}, {Gonz{\'a}lez
  Hern{\'a}ndez}, {Bignamini}, {Claudi}, {Covino}, {Damasso}, {Desidera},
  {Giacobbe}, {Gonz{\'a}lez-{\'A}lvarez}, {Herrero}, {Leto}, {Maggio},
  {Molinari}, {Morales}, {Pagano}, {Petralia}, {Piotto}, {Poretti}, {Rebolo},
  {Scandariato}, {Su{\'a}rez Mascare{\~n}o}, {Toledo-Padr{\'o}n}, \& {Zanmar
  S{\'a}nchez}}]{Pinamontietal2022}
{Pinamonti}, M., {Sozzetti}, A., {Maldonado}, J., {et~al.} 2022, \aap, 664, A65

\bibitem[{{Raghavan} {et~al.}(2010){Raghavan}, {McAlister}, {Henry}, {Latham},
  {Marcy}, {Mason}, {Gies}, {White}, \& {ten Brummelaar}}]{raghavan2010}
{Raghavan}, D., {McAlister}, H.~A., {Henry}, T.~J., {et~al.} 2010, \apjs, 190,
  1

\bibitem[{{Rajpaul} {et~al.}(2015){Rajpaul}, {Aigrain}, {Osborne}, {Reece}, \&
  {Roberts}}]{rajpaul2015}
{Rajpaul}, V., {Aigrain}, S., {Osborne}, M.~A., {Reece}, S., \& {Roberts}, S.
  2015, \mnras, 452, 2269

\bibitem[{{Ram{\'\i}rez} {et~al.}(2015){Ram{\'\i}rez}, {Khanal}, {Aleo},
  {Sobotka}, {Liu}, {Casagrande}, {Mel{\'e}ndez}, {Yong}, {Lambert}, \&
  {Asplund}}]{ramirez2015}
{Ram{\'\i}rez}, I., {Khanal}, S., {Aleo}, P., {et~al.} 2015, \apj, 808, 13

\bibitem[{{Ricker} {et~al.}(2015){Ricker}, {Winn}, {Vanderspek}, {Latham},
  {Bakos}, {Bean}, {Berta-Thompson}, {Brown}, {Buchhave}, {Butler}, {Butler},
  {Chaplin}, {Charbonneau}, {Christensen-Dalsgaard}, {Clampin}, {Deming},
  {Doty}, {De Lee}, {Dressing}, {Dunham}, {Endl}, {Fressin}, {Ge}, {Henning},
  {Holman}, {Howard}, {Ida}, {Jenkins}, {Jernigan}, {Johnson}, {Kaltenegger},
  {Kawai}, {Kjeldsen}, {Laughlin}, {Levine}, {Lin}, {Lissauer}, {MacQueen},
  {Marcy}, {McCullough}, {Morton}, {Narita}, {Paegert}, {Palle}, {Pepe},
  {Pepper}, {Quirrenbach}, {Rinehart}, {Sasselov}, {Sato}, {Seager},
  {Sozzetti}, {Stassun}, {Sullivan}, {Szentgyorgyi}, {Torres}, {Udry}, \&
  {Villasenor}}]{ricker2015}
{Ricker}, G.~R., {Winn}, J.~N., {Vanderspek}, R., {et~al.} 2015, Journal of
  Astronomical Telescopes, Instruments, and Systems, 1, 014003

\bibitem[{{Rossiter}(1924)}]{rossiter1924}
{Rossiter}, R.~A. 1924, \apj, 60, 15

\bibitem[{Schwarz(1978)}]{schwarz1978}
Schwarz, G. 1978, The Annals of Statistics, 6, 461

\bibitem[{{Shi} {et~al.}(2019){Shi}, {Zhu}, {Liu}, {Liu}, {Ding}, \&
  {Cheng}}]{shi2019}
{Shi}, Y.~Y., {Zhu}, Z., {Liu}, N., {et~al.} 2019, \aj, 157, 222

\bibitem[{{Sing} {et~al.}(2011){Sing}, {D{\'e}sert}, {Fortney}, {Lecavelier Des
  Etangs}, {Ballester}, {Cepa}, {Ehrenreich}, {L{\'o}pez-Morales}, {Pont},
  {Shabram}, \& {Vidal-Madjar}}]{sing2011}
{Sing}, D.~K., {D{\'e}sert}, J.~M., {Fortney}, J.~J., {et~al.} 2011, \aap, 527,
  A73

\bibitem[{{Sing} {et~al.}(2012){Sing}, {Huitson}, {Lopez-Morales}, {Pont},
  {D{\'e}sert}, {Ehrenreich}, {Wilson}, {Ballester}, {Fortney}, {Lecavelier des
  Etangs}, \& {Vidal-Madjar}}]{sing2012}
{Sing}, D.~K., {Huitson}, C.~M., {Lopez-Morales}, M., {et~al.} 2012, \mnras,
  426, 1663

\bibitem[{{Storn} \& {Price}(1997)}]{storn1997}
{Storn}, R. \& {Price}, K. 1997, Journal of Global Optimization

\bibitem[{{Teske} {et~al.}(2015){Teske}, {Ghezzi}, {Cunha}, {Smith}, {Schuler},
  \& {Bergemann}}]{teske2015}
{Teske}, J.~K., {Ghezzi}, L., {Cunha}, K., {et~al.} 2015, \apjl, 801, L10

\bibitem[{{Udry} {et~al.}(2019){Udry}, {Dumusque}, {Lovis}, {S{\'e}gransan},
  {Diaz}, {Benz}, {Bouchy}, {Coffinet}, {Lo Curto}, {Mayor}, {Mordasini},
  {Motalebi}, {Pepe}, {Queloz}, {Santos}, {Wyttenbach}, {Alonso}, {Collier
  Cameron}, {Deleuil}, {Figueira}, {Gillon}, {Moutou}, {Pollacco}, \&
  {Pompei}}]{udry2019}
{Udry}, S., {Dumusque}, X., {Lovis}, C., {et~al.} 2019, \aap, 622, A37

\bibitem[{{Wasson} \& {Kallemeyn}(1988)}]{WassonKallemeyn1988}
{Wasson}, J.~T. \& {Kallemeyn}, G.~W. 1988, Philosophical Transactions of the
  Royal Society of London Series A, 325, 535

\bibitem[{{Yana Galarza} {et~al.}(2016){Yana Galarza}, {Mel{\'e}ndez},
  {Ram{\'\i}rez}, {Yong}, {Karakas}, {Asplund}, \& {Liu}}]{yanagalarzaetal2016}
{Yana Galarza}, J., {Mel{\'e}ndez}, J., {Ram{\'\i}rez}, I., {et~al.} 2016,
  \aap, 589, A17

\bibitem[{{Yi} {et~al.}(2001){Yi}, {Demarque}, {Kim}, {Lee}, {Ree}, {Lejeune},
  \& {Barnes}}]{Yietal2001}
{Yi}, S., {Demarque}, P., {Kim}, Y.-C., {et~al.} 2001, \apjs, 136, 417

\bibitem[{{Zechmeister} \& {K{\"u}rster}(2009)}]{gls}
{Zechmeister}, M. \& {K{\"u}rster}, M. 2009, \aap, 496, 577

\bibitem[{{Zellem} {et~al.}(2015){Zellem}, {Griffith}, {Pearson}, {Turner},
  {Henry}, {Williamson}, {Ryleigh Fitzpatrick}, {Teske}, \&
  {Biddle}}]{zellem2015}
{Zellem}, R.~T., {Griffith}, C.~A., {Pearson}, K.~A., {et~al.} 2015, \apj, 810,
  11

\end{thebibliography}
\bibliographystyle{aa.bst} 


\end{document}